\documentclass[screen,sigconf]{acmart}

\usepackage[T1]{fontenc}
\usepackage[utf8]{inputenc}

\setcopyright{rightsretained}
\copyrightyear{2022}
\acmYear{2022}
\acmDOI{10.1145/3536221.3558058}

\acmConference[ICMI '22]{INTERNATIONAL CONFERENCE ON MULTIMODAL INTERACTION}{November 7--11, 2022}{Bengaluru, India}
\acmBooktitle{INTERNATIONAL CONFERENCE ON MULTIMODAL INTERACTION (ICMI '22), November 7--11, 2022, Bengaluru, India}
\acmISBN{978-1-4503-9390-4/22/11}

\usepackage{enumitem} % Better control of lists and their spacing

\setlist{nolistsep} % Remove horizontal spacing in lists
\usepackage{caption}
\usepackage{subcaption}
\usepackage{pifont}
\newcommand{\cmark}{\ding{51}}

\usepackage{rotating}

\usepackage[normalem]{ulem} % To allow strikeout text

\begin{document}

\title[The GENEA Challenge 2022]{The GENEA Challenge 2022: \\A large evaluation of data-driven co-speech gesture generation}
% \\ Benchmarking the State of the Art


\author{Youngwoo Yoon}

\authornote{Equal contribution and joint first authors.}
\email{youngwoo@etri.re.kr}
\affiliation{%
  \institution{ETRI}
  \city{Daejeon}
  \country{Republic of Korea}
}

\author{Pieter Wolfert}
\authornotemark[1]
\email{pieter.wolfert@ugent.be}
\affiliation{%
 \institution{IDLab, Ghent University – imec}
 \city{Ghent}
 \country{Belgium}}
 
\author{Taras Kucherenko}
\email{tkucherenko@ea.com}
\authornotemark[1]
\affiliation{
  \institution{SEED -- Electronic Arts (EA)}
  \city{Stockholm}
  \country{Sweden}}
 
\author{Carla Viegas}
\email{cviegas@andrew.cmu.edu}
\affiliation{%
 \institution{Carnegie Mellon University}
 \city{Pittsburgh}
 \country{USA}}
\affiliation{%
 \institution{%
 NOVA University Lisbon}
 \city{Lisbon}
 \country{Portugal}}
 
\author{Teodor Nikolov}
\email{tnikolov@hotmail.com}
\affiliation{%
 \institution{Umeå University}
 \city{Umeå}
 \country{Sweden}
 }

\author{Mihail Tsakov}
\email{tsakovm@gmail.com}
\affiliation{%
 \institution{Umeå University}
 \city{Umeå}
 \country{Sweden}
 }

\author{Gustav Eje Henter}
\email{ghe@kth.se}
\affiliation{%
  \institution{KTH Royal Institute of Technology}
  \city{Stockholm}
  \country{Sweden}
}

\renewcommand{\shortauthors}{Yoon, Wolfert, Kucherenko et al.}

\begin{abstract}
% IUI abstract draft:
%Co-speech gestures, gestures that accompany speech, play an important role in human communication. Automatic co-speech gesture generation is thus a key enabling technology for embodied conversational agents (ECAs). Research into gesture generation is rapidly gravitating towards data-driven methods. Unfortunately, individual research efforts in the field are difficult to compare: there are no established benchmarks, and each study tends to use its own dataset, motion visualisation, and evaluation methodology. To address this situation, we launched the GENEA Challenges, a series of gesture-generation challenges wherein participating teams built automatic gesture-generation systems on a common dataset, and the resulting systems were evaluated in parallel in a large, crowdsourced user study using the same motion-rendering pipeline. Since differences in evaluation outcomes between systems are solely attributable to differences between the motion-generation methods, this enables benchmarking recent approaches against one another in order to get a better impression of the state of the art in the field. This paper reports on the purpose, design, results, and implications of the second GENEA challenge. 
%Co-speech gesture generation is a key enabling technology for embodied conversational agents.
This paper reports on the second GENEA Challenge to benchmark data-driven automatic co-speech gesture generation. Participating teams used the same speech and motion dataset to build gesture-generation systems. Motion generated by all these systems was rendered to video using a standardised visualisation pipeline and evaluated in several large, crowdsourced user studies. Unlike when comparing different research papers, differences in results are here only due to differences between methods, enabling direct comparison between systems. This year's dataset was based on 18 hours of full-body motion capture, including fingers, of different persons engaging in dyadic conversation. Ten teams participated in the challenge across two tiers: full-body and upper-body gesticulation. For each tier we evaluated both the human-likeness of the gesture motion and its appropriateness for the specific speech signal. Our evaluations decouple human-likeness from gesture appropriateness, which previously was a major challenge in the field.

The evaluation results are a revolution, and a revelation. Some synthetic conditions are rated as significantly more human-like than human motion capture. To the best of our knowledge, this has never been shown before on a high-fidelity avatar. On the other hand, all synthetic motion is found to be vastly less appropriate for the speech than the original motion-capture recordings.
%The main highlight of the results is synthetic motion that is rated as significantly more human-like than the motion from the motion-capture recordings. To the best of our knowledge this has never been shown before on a high-fidelity avatar. We have also managed to decouple human-likeness from gesture appropriateness in our evaluations, which previously was a major challenge in the field.

%
% Alternative intro and ending for the GENEA Workshop paper abstract:
%Automatic gesture generation is a field of growing interest, and a key technology for enabling embodied conversational agents.
%(...)
%This paper reports on the purpose, design, and of our challenge, with each individual team's entry described in a separate paper also presented at the GENEA Workshop.
%

\end{abstract}



%% the work being presented. Separate the keywords with commas.I don't think arXiv offers any way to collect articles into proceedings.

\settopmatter{printfolios=true} % Turn this on for page numbering on arXiv

\maketitle

\section{Introduction}
% What writing conventions do we use? UK spelling with serial comma? Hyphenation patterns.
This paper is concerned with systems for automatic generation of nonverbal behaviour, and how these can be compared in a fair and systematic way in order to advance the state-of-the-art.
This is of importance as nonverbal behaviour plays a key role in conveying a message in human communication \cite{mcneill1992hand}.
A large part of nonverbal behaviour consists of so called co-speech gestures, spontaneous hand and body gestures that relate closely to the content of the speech \cite{bergmann2011relation}, and that have been shown to improve understanding \cite{holler2018processing}.
Embodied conversational agents (ECAs) benefit from gesticulation, as it improves interaction with social robots \cite{salem2011friendly} and willingness to cooperate with an ECA \cite{salem2013err}. 
%Knowledge of how and when to gesture is also needed. This can for example be learned from interaction data; see, e.g., \cite{jonell2020let} and references therein.

Synthetic gestures used to be based on rule-based systems, e.g., \cite{cassell2001beat,salvi2009synface}; see \cite{wagner2014gesture} for a review. These are gradually being supplanted by data-driven approaches, e.g., \cite{bergmann2009GNetIc,levine2010gesture,chiu2015predicting,kucherenko2021speech2properties2gestures}, with recent work \cite{yoon2019robots,kucherenko2020gesticulator,alexanderson2020style,yoon2020speech} showing improvements in gesticulation production for ECAs.
However, results from different gesture-generation studies are typically not directly comparable \cite{wolfert2021review}.
Studies usually rely on different data sources to train their models. 
The visualisations of their generated gestures often have different avatars and production values, which can affect the perception of the gestures.
%obscure the quality of the underlying gesture-generation approach.
On top of this, studies make use of a variety of different methodologies to evaluate the gestures. 
All these differences are, however, external to the actual methods that drive the gesture generation. 
%By providing a common dataset for building gesture-generation systems, and common evaluation standards with a common visualisation procedure, any differences between different syste.
%one can control for these sources of variation, and enable direct comparison between different methods for co-speech gesture generation.

In this paper, we report on the GENEA Challenge 2022,
%\footnote{GENEA stands for ``Generation and Evaluation of Non-verbal Behaviour for Embodied Agents''}
the second joint gesture-generation challenge.
(GENEA stands for ``Generation and Evaluation of Non-verbal Behaviour for Embodied Agents''.)
The aim of the challenge is not to select the best team -- it is not a contest, nor a competition -- but to be able to directly compare different approaches and outcomes.
By providing a common dataset for building gesture-generation systems, along with common evaluation standards and a shared visualisation procedure, we control for all other sources of variation except the system building itself.
This makes it possible to assess and advance the state of the art in gesture generation, and to measure the gap between it and natural co-speech gestures.
Comparing different methods and their performance also helps identify what matters most in gesture generation, and where the bottlenecks are.
In particular, this year's results make it abundantly clear that natural-looking data-driven gesture motion is achievable today, but that synthetic gestures are much less appropriate for the accompanying speech than the ground-truth motion is.
%Challenge participants benefit by working on the same problem together with researchers interested in the same topic, strengthening the research community, and get an opportunity to compare their systems to other competitive systems in a large and carefully-executed joint evaluation. They also get the chance to publish their work at ACM ICMI
%
%This, and help establish where the state of the art is, and what.
% which could lead to new standardised evaluation and generation methods.
% identify and advance the state of the art
% assess the gap between natural and synthetic gestures
% identify what matters most in generating good gestures, in various aspects
% enable system owners and creators to get good feedback on the performance of theirs systems (no fee)
% bringing together researchers interested in the same task to work on the same problem
% creating data that enables experimentation and development of objective and subjective evaluation metrics for gesture generation
% future benchmarking using the same setup and code
%
%todo: what did the blizzard challenge achieve? specific info on blizzard challenge is not relevant perse (except the meta stuff)
%Unique for this field is the cross-comparison of different systems by different researchers on one and the same dataset. 
%
%Participants are required to write down their methods, results and experience in a system paper, which is presented at the workshop. 
%
Our concrete contributions are:
\begin{enumerate}
    \item Four large-scale user studies that jointly evaluate a large number of gesture-generation models on a common dataset using a common 3D model and rendering method.
    \item Demonstrating a new method for subjective assessment of gesture appropriateness for speech, that successfully controls for the human-likeness of the motion.
    %\item Four large-scale user studies assessing the human-likeness and appropriateness of submitted motion.
    \item To the best of our knowledge, the first results that identify synthetic gesture motion that surpasses the human-likeness of good motion capture data on a high-fidelity avatar.
    \item The first clear evidence that synthetic gestures are much less appropriate for the specific speech than natural motion is, even when controlling for the human-likeness of the motion.
    \item Providing open code and high-quality,
    %data in the spirit of open source and reproducible research,
    to facilitate reproducibility and
    enable future research to compare and benchmark against systems from the challenge.
    %This includes pre-processed multimodal training, validation, and test datasets; the standardised visualisation; a large number of subjective responses from the studies; and evaluation and analysis code.
    \item Bringing researchers together in order to advance the state-of-the-art in gesture generation. %, and enabling future research to compare and benchmark against systems from the challenge.
\end{enumerate}
%In the long term, we anticipate further benefits in that systems, motion stimuli, and evaluation methods from the challenge can be used as benchmarks in future studies, and in starting to build a database of thoroughly-assessed motion stimuli for further research into objective and subjective.

The remainder of this paper first briefly discusses current gesture-evaluation practices and how challenges can help.
%their shortcomings, and the use of challenges in other fields.
We then describe this year's challenge data, setup, evaluation, and results, as well as implications of our findings.
Additional material is available via the project website at \href{https://youngwoo-yoon.github.io/GENEAchallenge2022/}{youngwoo-yoon.github.io/GENEAchallenge2022/}.
%for future challenges and gesture generation as a whole. 

%First, we provide an overview on relevant work in relation to benchmarks and challenges.  Section 3 covers the details of the challenge, the task description and the data set description. Section 4 describes the systems and teams, followed by section 5 that describes the methodology of our crowd sourced evaluation study. In section 6 we present both the objective and subjective results of the challenge evaluations, which we discuss in section 7. 

\section{Related work}
Most previous work proposing new gesture-generation methods incorporates an evaluation to support the merits of their method. Human gesture perception is highly subjective, and there are currently no widely accepted objective measures of gesture perception,
%due to the highly subjective aspect of human gestures,
so many publications have conducted human assessments instead.
However, previous subjective evaluations have several drawbacks, as reviewed in \cite{wolfert2021review}.
Some major issues are the coverage of systems being compared and the scale of the studies.
%Like in \cite{sadoughi2019speech,kucherenko2020gesticulator,kucherenko2021moving,alexanderson2020style}, proposed models are at most compared to one or two prior approaches (often a highly similar baseline) or possibly only to ablated versions of the same model.
%Proposed models were compared only with a few previous or ablated models \cite{sadoughi2019speech,kucherenko2020gesticulator, kucherenko2019analyzing, alexanderson2020style}, although
%A large number of studies do not compare their outcomes with other methods at all.
%, let alone other systems trained on the same data.
This creates an insular landscape where particular model families only are applied to particular datasets, and never contrasted against one another.
%As for scale, large evaluations are expensive, and studies may not be able to recruit enough participants, thus leaving the differences between many pairs of studied systems unresolved and not statistically significant (cf.\ \cite{yoon2019robots,yoon2020speech}).
%In terms of the study scale, Yoon et al.\ failed to show statistical significance for the majority of pairs of compared systems, due to the low number of evaluation participants.
%Questionnaires, which are one popular evaluation methodology (cf.\ \cite{salem2012generation,ishi2018speech,bergmann2010individualized}) demand a lot of time and cognitive effort even before scaling up.
%may mot be be feasible to scale up at all due to the cognitive effort they demand of test participants.
%In addition, the items used in questionnaires differs across studies and the set of questions used is often not standardised.
Evaluations also sometimes fail to anchor system performance against natural (``ground truth'') motion from test data held out from training.
%database, e.g., \cite{salem2012generation,ishii2018generating,le2012evaluating}.
%although systems often do compare their system's output to the ground truth from their data, many fail to do so \cite{salem2012generation, ishii2018generating, le2012evaluating}. 
%Using questionnaires to assess the quality of generated motion is very popular, but questionnaires come at a cost, since they require more cognitive effort to complete, yet they are often used in user-studies \cite{yoon2019robots, salem2012generation, ishii2018generating, ishi2018speech, bergmann2010individualized,  shimazu2018generation}. 
%Another significant difference between
Studies also differ in how the motion is visualised, where some prior work
%displaying motion through stick figures
%(e.g., \cite{wolfert2019should,kucherenko2019analyzing})
displays motion through stick figures, or applies it to a physical agent.
%(e.g., \cite{salem2012generation,ishi2018speech}).
Neither of these may allow the same expressiveness or range of motion as a high-quality 3D-rendered avatar.
%in, e.g., \cite{alexanderson2020style,kucherenko2020gesticulator}.

%cite more evaluation papers , with grouping them on 'errors' 

%Also, implementation tweaks were inevitable to train different models on the same dataset, and it may cause fairness issues, unless there is supervision by the original authors.
%Wolfert et al.\ \cite{wolfert2019should} conducted a benchmarking user study for beat gestures. They compared the data-driven gesture generation method \cite{kucherenko2019analyzing} and manually crafted beat gestures. 
%However, only one gesture generation model using speech audio context was used and the video stimuli were realised as stick figures which made it difficult to assess the gestures.

Other fields have done well using challenges to standardise evaluation techniques, establish benchmarks, and track and evolve the state of the art. 
For example, the Blizzard Challenges have since their inception in 2005 (see \cite{black2005blizzard}) helped advance our sister field of text-to-speech (TTS) technology and identified important trends in the specific strengths and weaknesses in different speech-synthesis paradigms \cite{king2014measuring}. 
%These challenges are open to both academia and industry.
%Participants are provided a common dataset of speech audio and associated text transcriptions, and use these to build a synthetic voice. 
%The resulting voices are then evaluated in a large, joint evaluation.
Data, evaluation stimuli, and subjective ratings remain available after these challenges, and have been widely used both for benchmarking subsequent TTS systems, e.g., \cite{szekely2012evaluating,charfuelan2013expressive}, and in research on the perception of natural and artificial speech, e.g., \cite{moller2010comparison,yoshimura2016hierarchical,mittag2020deep,saratxaga2016synthetic,govender2019using}.
%This challenge is defined by the use of common data and its open participation.
%After participants have submitted their systems, a common evaluation is done, the results of which are provided to the teams.
%Submitted entries are identified by anonymised labels in Blizzard challenge results, but in practice the vast majority of teams identify which label represents their entry in their paper at the Blizzard Challenge Workshop describing their submitted system.
%yet to the public it is not announced which system is from which team. 
%This has lead to the development of new and novel methods, driven by past results, and since participants had access to the same data, great steps have been made.

In 2020 we organised the first gesture-generation challenge, the GENEA Challenge 2020 \cite{kucherenko2021large}. In addition to being an exercise in benchmarking both new \cite{jinhong_lu_2020_4088376,vladislav_korzun_2020_4088609,thangthai2021speech} and previously-published \cite{alexanderson2020style, kucherenko2019analyzing, yoon2019robots} gesture-generation methods, the results of that challenge have since helped improve gesture-generation benchmarking in other ways as well. Researchers have, for example, used the 2020 visualisation \cite{wang2021integrated}, and the objective \cite{bhattacharya2021speech2affectivegestures} and subjective \cite{yoon2021sgtoolkit} evaluation methodologies, as a basis for future research. The data has also been used to benchmark subsequent gesture-generation models \cite{ferstl2021expressgesture, yazdian2021gesture2vec}, and even for automatic quality assessment \cite{he2022automatic}.
In this paper, we follow up on the 2020 challenge by reporting on the second gesture-generation challenge, the GENEA Challenge 2022.

\section{Task and data}
\label{sec:task_n_data}
%The GENEA Challenge 2022 focused on automatic of data-driven co-speech gestures.
%is to generate a sequence of 3D poses for a specific virtual character, that is to go together with a given recording of human speech.
%This is the same basic task as in the 2020 challenge, but.
%includes fingers and lower body is based on dyadic conversation with a multitude of different speakers
The GENEA Challenge 2022 focused on data-driven automatic co-speech gesture generation.
Specifically, given a sequence $\boldsymbol{s}$ of input features that describe human speech -- which could involve any combination of an audio waveform, a time-aligned text transcription, and a speaker ID -- the task is to generate a corresponding sequence $\hat{\boldsymbol{g}}$ of 3D poses describing gesture motion that an agent might perform while uttering this speech (facial expression is not considered).
%To enable direct comparison of different data-driven gesture-generation methods, all methods evaluated in the challenge were trained on the same gesture-speech dataset and their motion visualised using the same virtual avatar and rendering pipeline.
This is the same basic task as in the 2020 challenge, while at the same time we changed the dataset (as described below) and refined the evaluation (as detailed in Section\ \ref{sec:evaluation}).

%The task we focus on with the gesture generation challenge was to compare recent data-driven gesture generation in a fair way. To do so we made sure that every system participating in the evaluation was trained on the same gesture-speech dataset and was visualised on the same virtual avatar.

%\subsection{Dataset}
Compared to 2020, we wanted to expand the dataset to include finger motion, lower-body motion, and material from multiple speakers in dyadic interactions.
%We also wanted to consider data from dyadic interactions, which may provide more natural and interesting gestures than the Trinity Speech-Gesture Dataset \cite{ferstl2018investigating} used in 2020.
We therefore based our new challenge on the Talking With Hands 16.2M gesture dataset \cite{lee2019talking}, which comprises 50 hours of audio (close-talking microphones) and motion-capture recordings of several pairs of people having a conversation freely on a variety of topics, recorded in distinct takes each about 10 minutes long.
This is one of the largest datasets of parallel speech and 3D motion (in joint-angle space) publicly available in the English language.
We removed parts of the dataset (46 out of 116 takes) that lacked audio or had low motion-capture quality, especially for the fingers.
Note that despite the dataset being dyadic by design, this year's challenge focused on generating one side of the conversation, without awareness of the interaction partner.
%other party in the interaction.
%(neither for the synthesis, nor for the evaluation).

Speech data was shared with participants as WAV audio with no additional processing beyond the anonymisation  applied by \cite{lee2019talking}, which replaced many proper nouns with silence. We also provided text transcriptions of the speech, in tab-separated value (TSV) files, and a metadata file with unique anonymous labels for each speaker.
%We did not apply any processing to the WAV audio in the data provided by \cite{lee2019talking}. note that many proper nouns had been removed and replaced with silence there, for anonymity.
%To obtain verbal information from the speech, we
The TSV files were created by first
%transcribing the audio recordings using 
applying \href{https://cloud.google.com/speech-to-text/}{Google Cloud automatic speech recognition},
%\footnote{\href{https://cloud.google.com/speech-to-text/}{cloud.google.com/speech-to-text/}},
followed by thorough manual review to correct recognition errors and add punctuation for all parts of the dataset (training, validation, and test).

Motion data was downsampled to 30 frames per second and further transformed in two ways.
Firstly, we updated the default skeletal definition relative to which all motion data is defined, from what appeared to be a contorted and arbitrary definition, to a standard ``T-pose''.
%The T-pose is an animation industry standard, whereby all joint rotation values are described in relation to a ``T''-shaped skeleton.
%This standard makes it easier to work with motion data, and it is often the case that 3D applications require this pose for transferring motion from one character onto another.
%, and it also has the benefit that joint rotation values are better distributed around the 0 value, keeping them further away from potential gimbal locks.
The data was recomputed to match this pose using motion re-targeting inside MotionBuilder, retaining as much of the original visual quality as possible, whilst ensuring that the data had no discontinuities (e.g., at rotations near $180^{\circ}$).
% next sentence can be deleted to shorten paragraph
We found that this transformation substantially improved the output of the baseline system UBA in Section \ref{ssec:teams}.
%, for which the process of tuning the hyper-parameters became easier,
%possibly due to values being more closely distributed around 0.
Secondly, we standardised the position and orientation of speakers in all takes.
Originally, each take would have the two speakers occupy two locations and face each other.
We standardised this on a per-take basis
%Our change makes it
such that all speakers, on average, face the same direction, and occupy the same location.
%More technically, in a right-hand $XYZ$ Cartesian coordinate system ($Y$-up, $Z$-forward), each speaker is on average positioned at world origin ($[X=0,Z=0]$), and on average facing positive Z-axis (a directional vector $[X=0,Y=0,Z=1]$).
%Averaging was done on a per-take basis, after taking 250 equidistant samples of the hips position and orientation, and using linear algebra operations to map the original values to the standardized ones.
This change was made to streamline data visualisation and to remove potential confusion due to different positions and orientations across different takes.
Motion data was shared with participants in the Biovision hierarchy (BVH) format.

The challenge data was split into a training set (18 h), a validation set (40 min), and a test set (40 min), with only the training and validation sets initially shared with the teams.
All these data subsets are publicly available via the Zenodo data release at \href{https://doi.org/10.5281/zenodo.6998230}{doi.org/10.5281/zenodo.6998230}.
The validation and test data each comprised 40 \emph{chunks} (contiguous excerpts approximately one minute long), to promote generation methods that are stable over long segments of speech, and was restricted to recordings (``takes'' in the nomenclature of \cite{lee2019talking}) with finger motion tracking for the chosen speaker.
%Some recordings with finger capture data were excluded from consideration based on poor motion-capture quality, based on visual inspection of a short sample from each recording.
The validation data was intended for internal benchmarking during development, so participants were allowed to train their final submitted models on both training and validation data if they wished.

Teams were allowed to only train on a subset of the data and were allowed to enhance the data they trained on however they liked.
%Teams were allowed to enhance the data that they trained on however they liked, %for instance by manual annotation, by post-processing the speech and the motion, and/or by only training on a subset of the data.
%and may choose to only train on a subset of the data.
They were also allowed to make use of additional speech data (audio and text) from other sources, and models derived from such data, e.g., BERT \cite{devlin2018bert} and Wav2Vec \cite{wav2vec2020}.
However, it was not permitted to use any other motion data, nor any pre-trained motion models, other than what we provided for the challenge.

\section{Setup and participation}
The challenge began on May 16, 2022, when speech-motion training data was released to participating teams.
Test inputs (WAV, TSV, and speaker ID, but no motion output) were released to the teams on June 20, with teams required to submit BVH files with their generated gesture motion for these inputs by June 27.
%Motion had to be submitted in the same %Biovision Hierarchy (
%BVH format used by the challenge training data.
Manual tweaking of test inputs or the output motion was not allowed, since the idea was to evaluate synthesis performance in an unattended setting.
%Prior to entering the evaluation,
As a precondition for participating in the evaluation, teams agreed to submit a companion paper describing their system for review and possible publication at ACM ICMI.
%Evaluations took place after after the generated motion was submitted.

\subsection{Tiers}
This year's challenge evaluation was divided into two tiers, one for full-body motion and one for upper-body motion only.
%Each tier has it's own reasons for being included.
%On the one hand, the data comprises recorded full-body motion from conversational interactions.
%It can furthermore be argued human embodied conversation uses the full body.
%Also, generating full-body behaviour seems like a harder problem, since it represents a higher-dimensional probability distribution which is harder to learn from a statistical perspective, so if full-body generation is solved, restricted versions of the problem can be expected to be solved as well.
%On the other hand, it is debatable to what extent the motion of the lower body whilst speaking constitutes co-speech gestures, that depend on the speech over other aspects such as stance in response to the other parties in a conversation (data which was not provided this year).
%Focusing on the upper body also is more consistent with earlier evaluations of co-speech gesture generation, such as the GENEA Challenge 2020 \cite{kucherenko2021large}.
%Because it is not clear which perspective to apply, this year's GENEA evaluation included a tier each for full-body and upper-body motion.
Teams could enter motion into either tier, or into both, but could only make one submission per tier.
Teams that entered into both tiers were allowed to submit different motion (BVH files) to each tier, if they wished.
Both tiers used the same training data but differed in which parts of the avatar that were allowed to move, and in the camera angle used for the video stimuli in the evaluation, as follows:
\begin{description}
\item[Full-body tier] In this tier, the entire virtual character was free to move, including moving around in space relative to the fixed camera.
Motion was visualised from an angle facing the character that showed most of the legs, but not where the feet touched the ground.
This perspective was chosen to show as much as possible of the character, whilst obscuring foot penetration or foot sliding artefacts from view, 
%so that such artefacts (which
since these artefacts arguably do not relate to co-speech gestures.
%would not needlessly influence the ratings.
For an example of this camera perspective, see Figure\ \ref{fig:hemvipgui}.
\item[Upper-body tier] In this tier, the virtual character used a fixed position and a fixed pose from the hips down, with only the upper body free to move.
Motion was visualised from a camera angle facing the character, cropped slightly below the hips, such that the hands always should remain in view.
Any motion of the lower-body joints in submitted BVH files was ignored by the visualisation.
This camera perspective is shown in Figure\ \ref{fig:evaluation_interface_pairwise}.
\end{description}

\subsection{Baselines and participating teams}
\label{ssec:teams}
The challenge evaluation featured three types of motion sources: natural motion capture from the speakers in the database, baseline systems based on open code, and submissions by teams participating in the challenge.
We call each source of motion in a tier a \emph{condition} (not a ``system'', since not all conditions represent motion synthesised by an artificial system).
Each condition was assigned a unique three-letter \emph{label} or \emph{condition ID}, where the first character signifies the tier, with F for the full-body tier and U for the upper-body tier.

Natural motion was labelled \textbf{FNA} in the full-body tier and \textbf{UNA} in the upper-body tier (NA for ``natural'').
These conditions can be seen as a top line, and surpassing their performance essentially means outperforming the dataset itself, subject to limitations due to the motion capture and visualisation.
%Ignoring limitations due to the motion capture and visualisation, these conditions can be seen as a top line, and surpassing natural motion in the evaluations will essentially entail a measure superhuman performance, with some caveats discussed in Section\ \ref{ssec:humlikecomments}.
\begin{table*}
%\small
\centering
\caption{Conditions participating in the evaluation. Teams are ordered alphabetically. The following non-standard abbreviations were used: \textit{AR} for ``Auto-regression'',
%\textit{CNN} for ``Convolutional Neural Network'', \textit{RNN} for ``Recurrent Neural Network'',
\textit{SA} for ``Neural self-attention'' (e.g., Transformers), \textit{GANs} for ``Generative adversarial networks or adversarial loss terms'',
%\textit{VAEs} for ``Variational auto-encoders'',
%\textit{H.\ rules} for ``Hand-crafted rules'',
and \textit{MM} for ``Motion matching'',
%The following column names were shortened:
\textit{Frame-wise} for ``Generating output frame-by-frame'', \textit{Stoch.\ output} for ``Stochastic output'',
%(different output possible even if the inputs are the same),
and \textit{Smoothed} for ``Smoothing was applied''.}
\small%
\begin{tabular}{@{}l|ccc|ccccl|ccc@{}}
\toprule 
Baseline or team name & \multicolumn{3}{c|}{Inputs used} & \multicolumn{5}{c|}{Techniques used} & Frame- & Stoch. & Smoo\tabularnewline
 & Aud. & Text & Sp.\ ID & AR & RNNs & SA & VAEs & Other & wise & output & thed\tabularnewline
%\midrule
%Natural motion & \cmark & \cmark & \cmark & \multicolumn{7}{c@{}}{Not relevant} \tabularnewline
\midrule 
Audio-only baseline \cite{kucherenko2019analyzing} & \cmark &  &  &  &  \cmark &  &  &  & \cmark &  & \cmark \tabularnewline
Text-only baseline \cite{yoon2019robots} &  & \cmark &  & \cmark & \cmark  &  &  &  & \cmark &  & \cmark \tabularnewline
\midrule 
DeepMotion \cite{lu2022deepmotion} & \cmark & \cmark &  & \cmark &  & \cmark & \cmark & CNNs & \cmark & \cmark & \tabularnewline
DSI \cite{saleh2022hybrid} & \cmark &  &  & \cmark & \cmark & \cmark &  &  &  &  & \tabularnewline
FineMotion \cite{korzun2022recell} & \cmark & \cmark &  & \cmark & \cmark &  &  &  & \cmark &  & \cmark\tabularnewline
Forgerons \cite{ghorbani2022exempler} & \cmark &  &  & \cmark & \cmark &  & \cmark &  & \cmark & \cmark &   \tabularnewline
GestureMaster \cite{zhou2022gesturenaster} & \cmark & \cmark & \cmark &  &  &  &  & Hand-crafted rules, MM &  &  & \cmark\tabularnewline
IVI Lab \cite{chang2022ivi} & \cmark & \cmark & \cmark & \cmark & \cmark &  &  &  & \cmark & \cmark & \cmark\tabularnewline
Murple AI lab & \multicolumn{11}{c@{}}{%\sout{\hspace{3cm}}
No paper submitted% \sout{\hspace{3cm}}
}
\tabularnewline
ReprGesture \cite{yang2022reprgesture} & \cmark & \cmark & \cmark & \cmark & \cmark & \cmark & \cmark & CNNs, GANs &  &  & \cmark\tabularnewline
TransGesture \cite{kaneko2022transgesture} & \cmark &  &  & \cmark & \cmark &  &  &  & \cmark &  & \cmark\tabularnewline
UEA Digital Humans \cite{windle2022uea} & \cmark & \cmark & \cmark &  & \cmark &  &  &  & \cmark &  & \tabularnewline
\bottomrule
\end{tabular}
\label{tab:conditions}
\end{table*}

The natural top line can be contrasted against the two baseline systems included in the challenge, which represent previously published gesture-generation approaches with free and open code, adapted to run on the 2022 challenge training data.
These two baselines were:
\begin{description}
\item [Text-based baseline (FBT/UBT)]
This motion was generated by the gesture-synthesis approach from \citep{yoon2019robots} (which takes text transcriptions with word-level timestamps as the input) but adapted to joint rotations as described in \cite{kucherenko2021large}.
%A neural sequence-to-sequence architecture is used, where an encoder processes a sequence of speech words and a decoder outputs a sequence of human poses.
Motion from this baseline used a fixed lower body but was included in both tiers, as conditions \textbf{FBT} and \textbf{UBT}
(B for ``baseline'' and T for ``text'').
The code is available at \href{https://github.com/youngwoo-yoon/Co-Speech_Gesture_Generation/}{github.com/youngwoo-yoon/Co-Speech\_Gesture\_Generation/}.
\item [Audio-based baseline (UBA)]
This motion was generated by the Audio2Repr2Pose motion-synthesis approach \cite{kucherenko2019analyzing}, which only takes speech audio into account when generating output, adapted to joint rotations as described in \cite{kucherenko2021large}.
%This model uses a chain of two neural networks: one maps from speech to pose representation and another decodes the representation to a pose, generating motion frame by frame by sliding a window over the speech input.
Motion from this baseline was only included in the upper-body tier, as condition \textbf{UBA} (A for ``audio'').
Code is available in the challenge GitHub repository at \href{https://github.com/genea-workshop/Speech_driven_gesture_generation_with_autoencoder/tree/GENEA_2022/}{github.com/genea-workshop}.
\end{description}
These are the same baselines as in the GENEA Challenge 2020.
They were included to track the progress of the field and to provide continuity between different years of the challenge.
%Both these baselines were also included in the GENEA Challenge 2020, and are intended to provide continuity between different years of the challenge, and also to track the progress of the field with respect to an unchanging baseline.

Separate from top lines and baselines, a total of 10 teams participated in the GENEA evaluation, with 8 \emph{entries} (a.k.a.\ \emph{submissions}) to the full-body tier and 8 entries to the upper-body tier.
Submissions were labelled with the prefix FS and US (S for ``submission'') depending on the tier, followed by a single character to distinguish between different submissions in the same tier.
In particular, challenge entries to the full-body tier were labelled \textbf{FSA}--\textbf{FSI}, and entries to the upper-body tier were labelled \textbf{USJ}--\textbf{USQ}.
Condition FSE was withdrawn before the evaluation.
These labels are anonymous and have no relationship to team names or identities, but teams are free to reveal their label(s) if they wish.
%Label FSE is not used, as that entry that was withdrawn before the start of the evaluation.

Table \ref{tab:conditions} lists the baselines and participating teams, with basic information about their approach and references to their system-description papers.
One team lacks information, since they did not submit a paper for review.

\section{Evaluation}
\label{sec:evaluation}
We conducted a large-scale, crowdsourced, joint evaluation of gesture motion from the 
10 full-body conditions and 11 upper-body conditions
%in Table \ref{tab:conditions}
%in parallel
using a within-subject design (i.e., every rater
%was exposed to and
evaluated all conditions in each tier).
%generated gesture motions submitted by the participating teams, along with some other conditions.
%The evaluation focused on gesture quality of the various submitted systems.
%The systems were evaluated in terms of the human-likeness of the gesture motion itself, as well as the appropriateness of the gestures for the given input speech.
%The central difference from other gesture-generation evaluations is that all systems in our evaluation used the same motion data, the same visualisation/embodiment, and were scored together using the same evaluation methodology; only the motion-generation systems differed between the different entries that were compared. This allows the performance of systems to be compared directly, and the design aspects that influence performance can be traced more efficiently than in most previous publications.
%Jonell \& Kucherenko et al.\ \cite{jonell2020iva_crowd} recently found that the results from crowdsourcing evaluations were not significantly different from in-lab evaluations in terms of results and consistency. We therefore adopted an entirely crowdsourced approach, as opposed to for example the Blizzard Challenge, which has used a mixed approach. Attention checks were used to exclude participants that were not paying attention, as detailed in Section\ \ref{subs:att_checks_hl}.
%
%\subsection{Evaluation structure}
%Although the challenge computes and provides some objective metrics of motion quality (see Section \ref{sec:objective}), the focus of the GENEA Challenge is on a large-scale, crowdsourced subjective evaluation of the generated gestures.
For each tier, two orthogonal aspects of the generated gestures were evaluated:
%, with one study per aspect and tier:
\begin{description}
\item[Human-likeness] Whether the motion of the virtual character looks like the motion of a real human, controlling for the effect of the speech. We sometimes use ``motion quality'' as a synonym for this.
\item[Appropriateness] (a.k.a.\ ``specificity'') Whether the motion of the virtual character is appropriate for the given speech, controlling for the human-likeness of the motion.
%We sometimes use ``specificity'' as a synonym for this.
\end{description}

% Seems useful for a journal
%Although an interesting question for a multispeaker dataset, we did not attempt to evaluate the specificity of the gesture motion style to different individuals in the database, since the data is too imbalanced to allow that.
%Other facets of appropriateness, such as emotional appropriateness, or the appropriateness of motion for a specific individual speaker, were not evaluated.
%More details about these evaluations are provided in Sections \ref{sec:humlike} and \ref{sec:approp} below, respectively.
%Speaker appropriateness , due to the data .
%A similar is true for other appropriateness aspects, such as emotion appropriateness.

%Although the speech and motion in the challenge comes from joint full-body motion capture of dyadic interactions with separate close-talking microphones for each speaker, this year's challenge only considered generating one side of the conversation, without awareness of the other party in the interaction (neither for the synthesis, not for the evaluation).

\subsection{Stimuli}
%The test set was deliberately made large to make it difficult to overfit to specific speech being evaluated.
%Like the GENEA Challenge 2020 and the Blizzard Challenges, not all test-set motion was included in the subjective evaluation.
From the 40 test-set chunks we selected 48 short \emph{segments} of test speech and corresponding test motion to be used in the subjective evaluations, based on the following criteria:
%Our rules for selecting these segments were as follows:
%\begin{itemize}
i) Segments should be around 8 to 10 seconds long.
%, and ideally not shorter than 6 seconds.
ii) The character should only be speaking, not passively listening, in the segments. %should only contain
%dialogue turns, or parts of turns, where the person in the recording is talking
(No turn-taking, but backchannels from the interlocutor were OK.)
iii) Segments should not contain any parts where \citeauthor{lee2019talking} had replaced the speech by silence for anonymisation.
%(This mostly affected proper nouns.)
iv) Segments should be more or less complete phrases, starting at the start of a word and ending at the end of a word, and not end on a ``cliffhanger''.
%\item Segments should end with some margin before the end of a chunk, to allow space for.
v) Finally, recorded motion capture in the segments (i.e., the FNA motion) should not contain any significant artefacts such as whole-body vibration or hands flicking open and closed due to poor finger tracking. This does not imply that the motion capture is perfect or completely natural,
%in all these segments evaluated, since the finger-tracking quality throughout the database does not allow our evaluations to reach that standard; it
just that the level of finger-tracking quality in the stimuli was consistent with the better parts of the source data.
%\end{itemize}
%: each segment should contain a complete phrase, there should be no censored audio parts in those segments, ...

The 48 segments selected in this way were between 5.6 and 12.1 seconds in duration and on average 9.5 seconds long.
%with the average duration being 9.5 seconds.
Audio was loudness normalised
%to $-$23 dB LUFS
following EBU R128 \cite{ebu2020loudness} to achieve a consistent listening volume in the user studies.
\begin{figure*}[!t]
\centering
%   \hfill
\begin{subfigure}[b]{0.416\textwidth}
    \includegraphics[width=\textwidth]{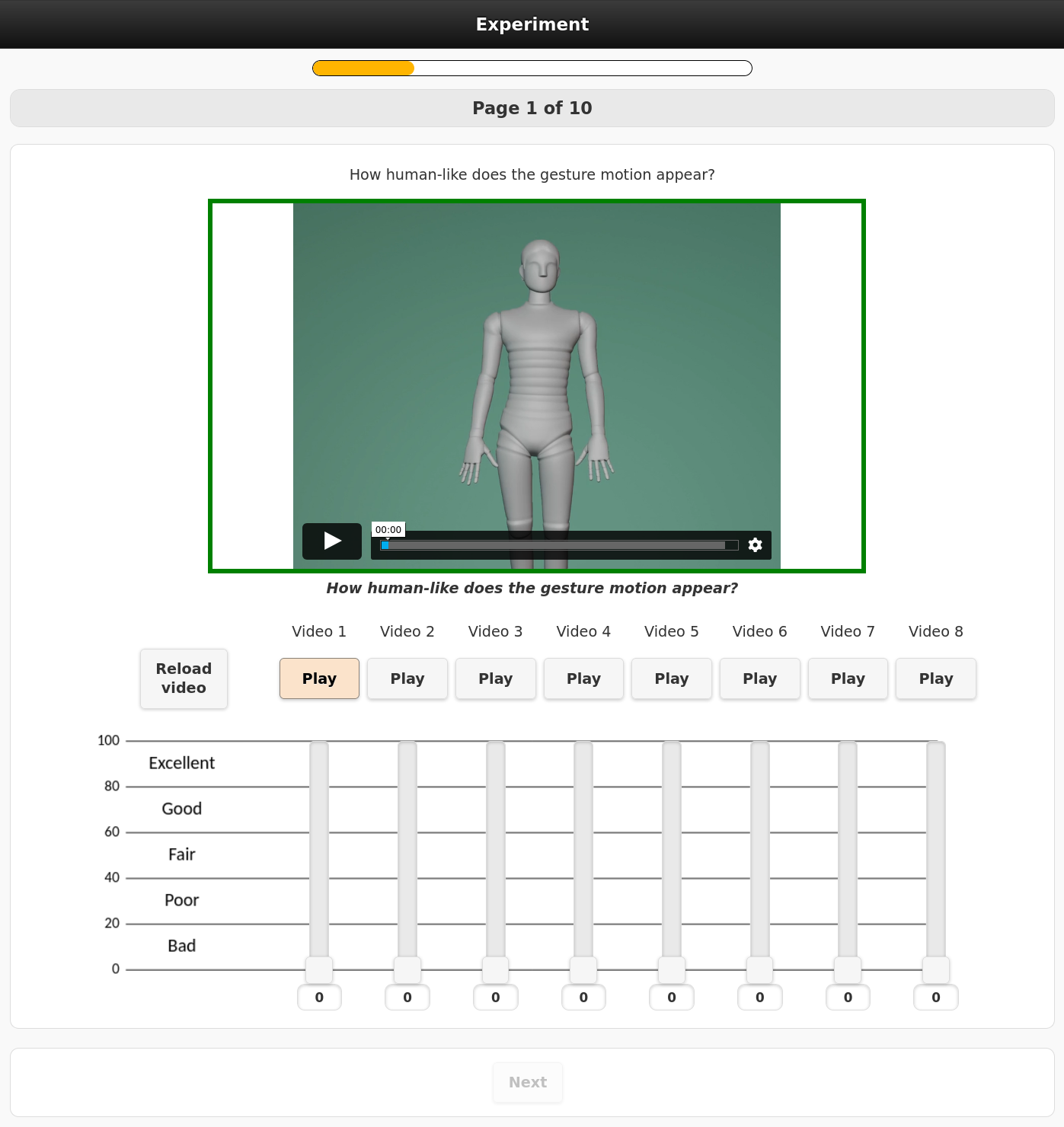}
    \caption{Human-likeness interface (HEMVIP) and full-body video}
    \label{fig:hemvipgui}
\end{subfigure}
\hfill\hfill
\begin{subfigure}[b]{0.564\textwidth}
    \centering\includegraphics[width=\columnwidth]{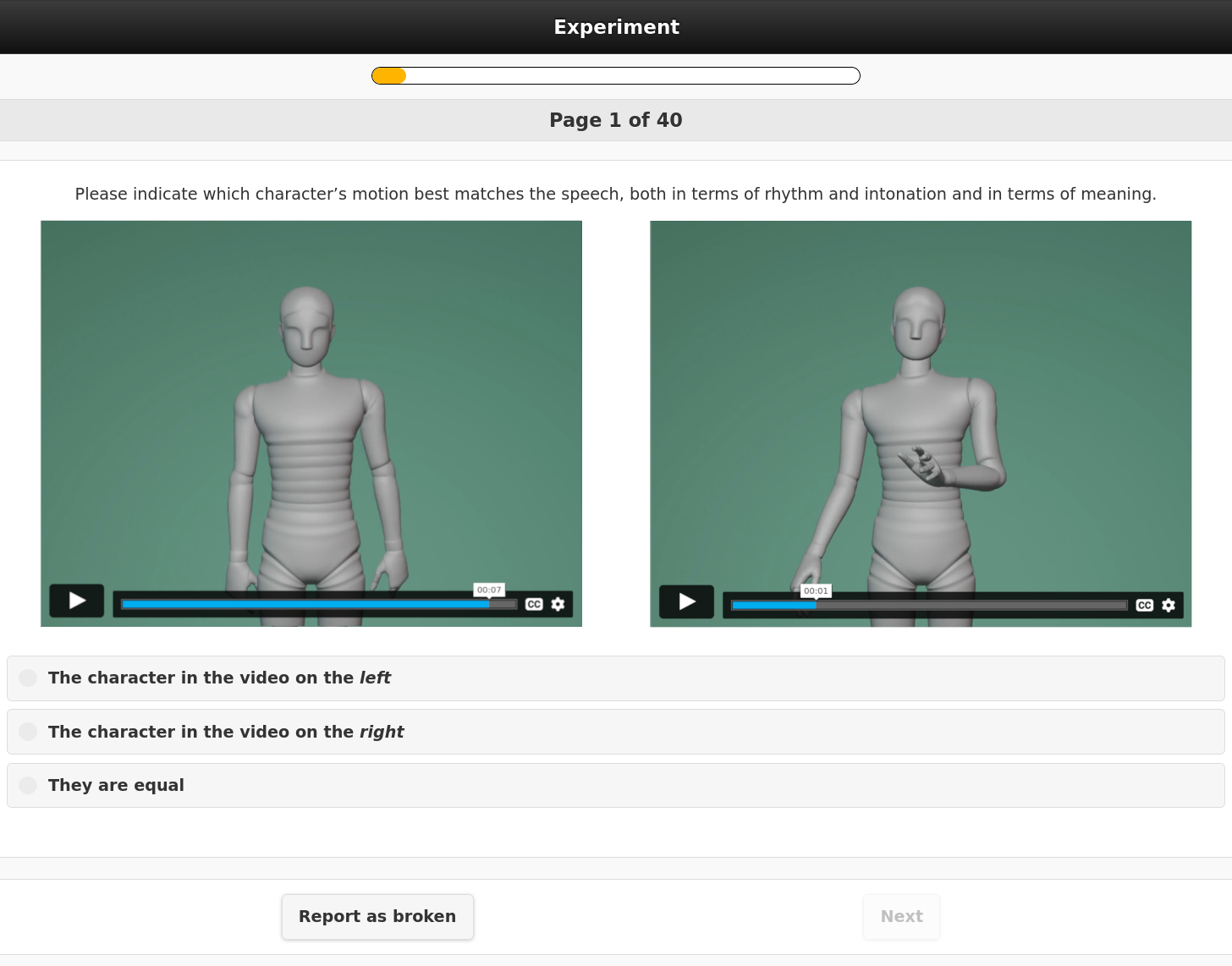}
    \caption{Appropriateness interface and upper-body videos}
    \label{fig:evaluation_interface_pairwise}
\end{subfigure}
%   \hfill
\vspace{-1.5pt}
\caption{Screenshots of the evaluation interfaces used in the studies, also showing the camera perspectives used by the tiers.}
%\vspace{-2.6ex}
\label{fig:interfaces}
\Description{The left figure shows the human-likeness interface. The interface presents a video showing a full-body avatar and eight sliders for rating. There are play buttons for each slider and one reload video button. The right figure shows the appropriateness interface. The interface presents two side-by-side videos showing an upper-body avatar. There is a radio option group for the question asking which character's motion best matched the speech. The options are 1) the character in the video on the left, 2) the character in the video on the right, 3) they are equal. The button 'Report as broken' is on the bottom.}
\vspace{-1.5pt}
\end{figure*}

%\subsubsection{Virtual avatar}
We used the same virtual avatar for all all videos rendered during the challenge and the evaluation. The avatar can be seen in Figure\ \ref{fig:interfaces}. The avatar originally had 56 joints (full body including fingers) and was designed to be gender neutral and omit eyes or mouth, to help evaluators focus on the rest of the body instead.
All teams had access to the official visualisation and rendering pipeline during the system-building phase, in the form of code, a portable Docker container, as well as a webserver to which BVH files could be submitted to be rendered as video.
%\subsubsection{Visualisation server}
%\label{subs:vis_ser}
%We also developed a visualisation server that enabled all participating teams to produce gesture-motion visualisations identical (except in resolution) to the video stimuli evaluated in the challenge. This was implemented using a Python-based web server which interfaced \href{https://www.blender.org}{Blender} 2.93.9.
%\footnote{\href{https://www.blender.org}{www.blender.org}}.
%Participants would send a 30-fps BVH file to the visualisation server, and these files were then processed as quickly as possible into videos visualising the motion on the avatar, in the order they came in.
%which then was put in a queue, to be processed into a video file order once a rendering worker became available.
%The same server was also used to render the final stimuli, but with the resolution increased to 960$\times$540 instead of 480$\times$270. (The lower resolution was used during the main part of the challenge to increase performance and throughput of the server, since 16 teams initially took part.)
%The input to the server was expected to be 20 fps.
The visualisation server code is provided at \href{https://github.com/TeoNikolov/genea_visualizer/}{github.com/TeoNikolov/genea\_visualizer/} and the rendered stimulus videos at \href{https://doi.org/10.5281/zenodo.6997925}{doi.org/10.5281/zenodo.6997925}.
%The code was available to the participants during the challenge, and they were free to use on their own servers. 

\subsection{Human-likeness evaluation}
\label{sec:humlike}

%\subsection{Evaluation procedure}
The human-likeness evaluation
%of this year's GENEA Challenge
closely followed the human-likeness evaluation in the GENEA Challenge 2020 \cite{kucherenko2021large}, by presenting multiple motion examples in parallel and asking the subject to provide a rating for each one.
%Specifically, it was based on the HEMVIP methodology \cite{jonell2021hemvip}, where multiple motion examples are presented in parallel and the subject is asked to provide a rating for each one.
All stimulus videos on the same page (a.k.a.\ screen) of the evaluation corresponded to the same speech segment but different conditions.
The advantage of this
%parallel methodology
method, called HEMVIP (Human Evaluation of Multiple Videos in Parallel) \cite{jonell2021hemvip},
is that differences in rating between the different conditions can be analysed using pairwise statistical tests, which helps control for variation between different subjects and different input speech segments; see \cite{jonell2021hemvip}.
The videos used in this evaluation had the audio removed, since it has been found that speech and gesture perception influence each other \cite{bosker2021beat} and can confound motion evaluations \cite{jonell2020let}.
%(see Section\ \ref{ssec:pairwisehumlike}), where each observation was gathered from the same person watching videos featuring the same speech segment -- the only difference between videos on the same page being which condition that generated the motion of the virtual character in each video clip.
%For a detailed explanation of the evaluation interface we refer the reader to \cite{jonell2021hemvip}, which introduced and validated the evaluation paradigm for gesture-motion stimuli.
Code is provided at \href{https://github.com/jonepatr/hemvip/tree/genea2022/}{github.com/jonepatr/hemvip/tree/genea2022/}.

%The HEMVIP evaluations presented 8 video stimuli to be rated per page.
Each evaluation page asked participants ``How human-like does the gesture motion appear?'' and presented eight video stimuli to be rated
%Study participants gave their ratings in response to this question
on a scale from 0 (worst) to 100 (best) by adjusting an individual GUI slider for each video.
An example of the evaluation interface can be seen in Figure\ \ref{fig:hemvipgui}.
%This screenshot is provided in the enclosed file \texttt{hemvip\_rating.png} under a Creative Commons Attribution 4.0 International (CC BY 4.0) license, for optional use in your papers.
%) will provided with future versions of this document, but they are overall very similar to the screenshots in \cite{jonell2021hemvip}.
Like in \cite{kucherenko2021large,jonell2021hemvip}, the 100-point rating scale was anchored by dividing it into successive 20-point intervals with labels (from best to worst) ``Excellent'', ``Good'', ``Fair'', ``Poor'', and ``Bad'',
%These labels were based on those associated with the 5-point scale used in
from the Mean Opinion Score ITU standard \cite{itu1996telephone}.
%for audio quality evaluation.
%Since it has been found that speech content can influence gesture perception and confound motion evaluations \cite{jonell2020let}, the videos seen by participants in these human-likeness evaluations (although they all corresponded to the same speech input and had the same length) were completely silent and did not include any audio.
%This way, ratings can only depend on the motion seen in the videos.

After reading the instructions, each subject completed one training page (to familiarise them with the task)
%and what the stimuli would look like)
followed by 10 pages of ratings for the evaluation.
Responses given on the training page were not included in the analysis.
The evaluation was balanced in exactly the same way as in \cite{kucherenko2021large}.
%The evaluation design was balanced such that each segment appeared on pages 1 through 10 with approximately equal frequency across all participants (segment order), and each condition was associated with each slider with approximately equal frequency across all pages (condition order).
%For any given participant and study, each of the 10 pages would use a different speech segment.
Condition FNA/UNA was included on every page
%Every page in the evaluation contained one stimulus video from condition FNA/UNA, visualising the motion-capture data gathered from the speaker at the same time that they uttered the speech in question.
to help calibrate evaluators' ratings and keep them consistent throughout.
%the test.
%assign a high numerical rating to the best motion,
Since motion-capture data projected onto a virtual character may not necessarily look
%be perceived as
perfectly natural, there was no requirement to rate the best motion as 100.
%Beyond FNA/UNA appearing on every page, the studies were balanced with the design goal that every combination of two distinct conditions should appear on the pages approximately equally often, and at least 600 times.
%
%The 10 test pages were preceded by a screen with instructions, and then by a designated training page showing a fixed set of videos with different motion, to familiarise participants with the task and what the stimuli would look like, before starting the study in earnest.
%After completing the rating pages, but before submitting the study, participants filled in a short questionnaire to gather broad, anonymous demographic information.
%The specific questions asked were similar to

\subsection{Appropriateness evaluation}
\label{sec:approp}
The appropriateness evaluation was designed to assess the link between the motion and the input speech, separate from the intrinsic human-likeness of the motion.
%In addition to human-likeness, the GENEA Challenge 2020 also evaluated the link between the motion and the input speech, by asking participants to assess how appropriate a given gesture was for the speech.
In the previous GENEA Challenge, appropriateness was evaluated using a HEMVIP-based rating study very similar to that for human-likeness, except that speech audio was included in the videos.
Test takers were asked to ignore the motion quality and only rate the appropriateness of the motion for the speech \cite{kucherenko2021large}.
Unfortunately, that evaluation was not altogether successful, since the \emph{mismatched} condition M -- which paired natural motion segments with unrelated speech segments, intended as a bottom line -- attained the second-highest appropriateness rating, above all synthetic systems.
This suggests a significant dependence between the
%perceived
human-likeness of a motion segment and its perceived appropriateness for speech, confounding the evaluation.
%This dependence acted as a confounding factor in the study, with the result that all systems ranked below natural-looking motion unrelated to the speech, intended as a bottom line in terms of appropriateness.

For the GENEA Challenge 2022, we decided to evaluate motion appropriateness for speech in a different way.
Our design goal
%for this year's challenge
was to assess appropriateness whilst controlling for the human-likeness of the motion in an effective way.
To do so, we took the idea of mismatching and used it within every condition:
%, and not just for the recorded motion-capture data FNA/UNA.
On each page, subjects were presented with a pair of videos containing the same speech audio.
Both videos contained motion from the same condition and thus had the same overall motion quality, but one was matched to the speech audio and the other mismatched, belonging to unrelated speech.
Whether the left or the right video was mismatched was randomised.
Subjects were then asked
%to pick the one video from the pair that best matched the speech.
%for each pair of videos, participants were asked
to ``Please indicate which character's motion best matches the speech, both in terms of rhythm and intonation and in terms of meaning.''
In response, they could choose the character on the left, on the right, or indicate that the two were equally well matched (``They are equal'', also referred to as \emph{equal}\ or a \emph{tie}).
We
%decided to ask
asked for preferences rather than ratings since there is evidence \cite{wolfert2021rate} that this is more efficient in pairwise comparisons like these.
A screenshot of the
%evaluation
interface used for the appropriateness studies is presented in Figure\ \ref{fig:evaluation_interface_pairwise}.

The extent to which test-takers prefer the character with the matched motion reveals how specific the gesture motion is to the given speech: Random motion will result in a 50--50 split, whereas conditions whose motion is more specifically appropriate to the input speech are expected to elicit a higher relative preference for the matched motion.
In this type of evaluation, condition M (the mismatched condition) from the 2020 challenge will perform at chance rate, rather than being tied for second highest as in 2020.
% as in the 2020 evaluation.
%an increasing preference for matched motion the more appropriate the motion in a condition is for the input speech.
%To our knowledge, t
This approach to control for motion quality was first piloted in \cite{jonell2020let}.

Concretely, we created the mismatched stimuli by taking the 48 existing speech and motion segments from the evaluation, and permuted the motion in between them such that no motion segment ever remained in its original place.
%Mathematicians call such a permutation a \emph{derangement}.
As the 48 different segments did not all have the same length, a longer or shorter segment of motion generally had to be excerpted from the motion chunks (original or generated), so as to match the new speech duration.
The starting point of the motion video was always the same as in the respective matched stimulus video (i.e., corresponding to the start of a phrase).
%The mismatched video stimuli are available at ...

After an instruction page and a training page, each subject evaluated 40 pages with one pair of videos each.
This means that subjects watched 80 videos total in each study, the same number of videos as was evaluated in the human-likeness studies (ignoring the training pages in all cases).
Each study was balanced such that each speech segment, condition, and order of the two videos appeared approximately equally many times.%
\begin{table*}[!t]
\centering%
\caption{Summary statistics of responses from all user studies, with 95\% confidence intervals. ``M.'' stands for ``matched'' and ``Mism.'' for ``mismatched''. ``Percent matched'' identifies how often subjects preferred matched over mismatched motion.}
\label{tab:stats}
%\hfill
\begin{subtable}[t]{0.49\textwidth}
\centering%
\caption{Full-body}%
\label{stab:fbstats}%
\small%
\begin{tabular}{@{}l|c|ccc|c@{}}
\toprule
& Median
& \multicolumn{4}{c@{}}{Appropriateness} \\
& human- & \multicolumn{3}{c|}{Num.\ responses}
& Percent matched\\
ID & likeness%Median rating
& M. & Tie & Mism. & (splitting ties)\\
\midrule
FNA & $70\hphantom{.0}\in{}[69,71]$
& 590 & 138 & 163 & $74.0\in{}[70.9,76.9]$\\
FBT & $27.5\in{}[25,30]$
& 278 & 362 & 250 & $51.6\in{}[48.2,55.0]$\\
FSA & $71\hphantom{.0}\in{}[70,73]$
& 393 & 216 & 269 & $57.1\in{}[53.7,60.4]$\\
FSB & $30\hphantom{.0}\in{}[28,31]$
& 397 & 163 & 330 & $53.8\in{}[50.4,57.1]$\\
FSC & $53\hphantom{.0}\in{}[51,55]$
& 347 & 237 & 295 & $53.0\in{}[49.5,56.3]$\\
FSD & $34\hphantom{.0}\in{}[32,36]$
& 329 & 256 & 302 & $51.5\in{}[48.1,54.9]$\\
FSF & $38\hphantom{.0}\in{}[35,40]$
& 388 & 130 & 359 & $51.7\in{}[48.2,55.1]$\\
FSG & $38\hphantom{.0}\in{}[35,40]$
& 406 & 184 & 319 & $54.8\in{}[51.4,58.1]$\\
FSH & $36\hphantom{.0}\in{}[33,38]$
& 445 & 166 & 262 & $60.5\in{}[57.1,63.8]$\\
FSI & $46\hphantom{.0}\in{}[45,48]$
& 403 & 178 & 312 & $55.1\in{}[51.7,58.4]$\\
\bottomrule
\end{tabular}
\end{subtable}%
\hfill\hfill
\begin{subtable}[t]{0.49\textwidth}
\centering%
\caption{Upper-body}%
\label{stab:ubstats}%
\small%
\begin{tabular}{@{}l|c|ccc|c@{}}
\toprule
& Median
& \multicolumn{4}{c@{}}{Appropriateness} \\
& human- & \multicolumn{3}{c|}{Num.\ responses}
& Percent matched\\
ID & likeness%Median rating
& M. & Tie & Mism. & (splitting ties)\\
\midrule
UNA & $63\hphantom{.0}\in{}[61,65]$
& 691 & 107 & 189 & $75.4\in{}[72.5,78.1]$\\
UBA & $33\hphantom{.0}\in{}[31,34]$
& 424 & 264 & 303 & $56.1\in{}[52.9,59.3]$\\
UBT & $36\hphantom{.0}\in{}[34,39]$
& 341 & 367 & 287 & $52.7\in{}[49.5,55.9]$\\
USJ & $53\hphantom{.0}\in{}[52,55]$
& 461 & 164 & 365 & $54.8\in{}[51.6,58.0]$\\
USK & $41\hphantom{.0}\in{}[40,44]$
& 454 & 185 & 353 & $55.1\in{}[51.9,58.3]$\\
USL & $22\hphantom{.0}\in{}[20,25]$
& 282 & 548 & 159 & $56.2\in{}[53.0,59.4]$\\
USM & $41\hphantom{.0}\in{}[40,42]$
& 503 & 175 & 328 & $58.7\in{}[55.5,61.8]$\\
USN & $44\hphantom{.0}\in{}[41,45]$
& 443 & 190 & 352 & $54.6\in{}[51.4,57.8]$\\
USO & $48\hphantom{.0}\in{}[47,50]$
& 439 & 209 & 335 & $55.3\in{}[52.1,58.5]$\\
USP & $29.5\in{}[28,31]$
& 440 & 180 & 376 & $53.2\in{}[50.0,56.4]$\\
USQ & $69\hphantom{.0}\in{}[68,70]$
& 504 & 182 & 310 & $59.7\in{}[56.6,62.9]$\\
\bottomrule
\end{tabular}
\end{subtable}%
%hfill
\end{table*}

\subsection{Test takers and attention checks}
\label{subs:att_checks_hl}
It has recently been found that crowdsourced evaluations are not significantly different from in-lab evaluations in terms of results and consistency \cite{jonell2020iva_crowd}.
The challenge therefore adopted an entirely crowdsourced approach.
%, as opposed to for example the Blizzard Challenge, which has used a mixed approach. Attention checks were used to exclude participants that were not paying attention, as detailed in Section\ \ref{subs:att_checks_hl}.
Test takers (a.k.a.\ subjects) were recruited through the crowdsourcing platform \href{https://www.prolific.co/}{Prolific}.
We used Prolific's built-in pre-screening tools to restrict the pool of test-takers in two ways: i) subjects were required to reside in any of six English-speaking countries, namely UK, IE, USA, CAN, AUS, and NZ, and ii) subjects were required to have English as their first language.

We conducted four user studies, two for human-likeness and two for appropriateness.
%, due to the two tiers.
A subject could take one or more studies, but could only participate in each study at most once, and could not use a phone or tablet to take the test.

Each study incorporated four attention checks per person, to make sure that subjects were paying attention to the task and remove insincere test-takers.
For the human-likeness studies, these attention checks took the form of a text message ``Attention! You must rate this video NN'' superimposed on the video.
``NN'' would be a number from 5 to 95, and the subject had to set the corresponding slider to the requested value, plus or minus 3, to pass that attention check.
%Which sliders on which pages that were used for attention checks was uniformly random, except that no page had more than one attention check, and the natural motion (condition FNA and UNA) was never replaced by an attention check.
For the appropriateness studies, the attention checks either displayed a brief text message over the gesticulating character, reading ``Attention! Please report this video as broken'', or they temporarily replaced the audio with a synthetic voice speaking the same message.
Subjects were exposed to two attention checks of each kind.
To pass the attention check, participants had to click a button marked ``Report as broken'' seen in Figure\ \ref{fig:evaluation_interface_pairwise}, forwarding them to the next pair of videos in the evaluation.
In all studies, the attention-check messages did not appear until a few seconds into each attention-check video, so that participants who only would watch the first seconds would be unlikely to pass the checks.

Subjects who failed two or more attention checks were removed from the respective study without being paid, since Prolific's policies do not allow rejecting a subject on the basis of a single failed attention check.
%Only the subjects who failed zero or one attention check for a study have been included in our analyses below.
%After watching and responding to all the videos but
Right before submitting their results, subjects also filled in a short questionnaire to gather broad, anonymous demographic information about the population taking the test.

%Responses to videos used for attention checks were not included in our analyses.

A design goal of the human-likeness studies was that every combination of two distinct conditions should appear on the pages approximately equally often, and at least 600 times (not counting FNA/UNA, which appeared on every page).
To meet this goal, we recruited 121 test takers that successfully passed the attention checks and completed the full-body study, and 150 test takers that successfully passed the attention checks and completed the upper-body study.
Of the 121 test takers in the full-body study, 60 identified as female, 60 as male, and 1 did not want to disclose their gender. 
%2 resided in AU, 2 in CA, 3 in IE, 110 in the UK, and 4 in the US. 
%Of the 150 test takers for the upper-body study, 74 identified as female, 75 as male and 1 did not want to disclose their gender. 
%1 resided in AU, 4 in IE, 134 in the UK, and 11 in the US.
The same numbers for the 150 upper-body test takers were 74, 75, and 1, respectively.
For the full-body test takers, 2 resided in AU, 2 in CAN, 3 in IE, 110 in the UK, and 4 in the USA.
The upper-body study had 1 AU, 4 IE, 134 UK, and 11 USA.

%(Since the latter study compared 11 conditions instead of only 10, it required more participants to reach the desired number of ratings pairs.)

% Seems useful for a journal
%All subjects passed all attention checks except, for a subject in the upper body study, who failed one attention check.

For the appropriateness studies, our design goal was for each condition to receive as many responses per condition as the number of ratings that each condition (aside from FNA/UNA)
%which were present on every page
received in the corresponding human-likeness evaluation.
This works out to 880 responses per condition in the full-body studies and 990 responses per condition in the upper-body studies.
Because a subject in these studies provided half as many responses as in a human-likeness study (40 vs.\ 80), the appropriateness studies needed to recruit approximately twice as many test takers.
%Since, ignoring attention checks, each subject in an appropriateness study provided half as many responses as in a human-likeness study (40 vs.\ 80), the appropriateness studies needed approximately twice the number of participants to reach their target.
In the end, 247 test takers successfully passed the attention checks in the full-body study, while 304 passed the attention checks in the upper-body study.
Of the 247 subjects in the full-body study, 137 identified as female, 107 as male, and 3 did not want to disclose their gender.
%3 were from Australia, 13 from Canada, 10 from Ireland, 2 from New Zealand, 211 from the UK and 8 from the USA. The mean age was 38 (SD 13.83). 
%Of the 304 participants in the upper body study, 127 identified as female, 173 as men and 4 did not want to disclose their gender.
%2 were from Australia, 10 from Canada, 1 from Ireland, 256 from the UK and 35 from the USA.
%The mean age was 38 (SD 12.84).
The same numbers for the 304 upper-body test takers were 127, 173, and 4, respectively.
For the full-body test takers, 3 resided in AU, 13 in CAN, 10 in IE, 2 in NZ, 211 in the UK, and 8 in the USA.
The upper-body study had 2 AU, 10 CAN, 1 IE, 256 UK, and 35 USA.

%All of these passed all attention checks, except for XX participants in the full-body study and XX participants in the upper-body study, who each failed one attention check.
%Each study participant contributed 36 ratings to the analyses after removing attention checks, unless they had to skip a page in the rare case of a video failing to load (which occurred approximately 1.6 times per 1000 videos presented).

Test takers were remunerated 6 GBP for each successfully completed human-likeness study.
Since the median completion time was 
%these studies took on average
28 minutes each, this corresponds to a median compensation just above 12 GBP per hour.
Similarly, the appropriateness studies took a median of 24 or 25 minutes to complete, and earned a reward of 5.5 GBP each, amounting to around 13 GBP per hour.
%, ``so that extreme completion times do not skew the data.'')
These compensation levels all exceed the UK national living wage.
%and also exceeds the highest living wage quoted by the Living Wage Foundation in the UK.
%(All numbers are as measured by Prolific, which uses the median rather than the mean for these calculations.)

%\subsection{Response data}
%121 participants successfully passed the attention checks and completed the full-body human-likeness study (one more than our explicit target of 120), while 150 participants successfully passed the attention checks and completed the upper-body human-likeness study.
%(Since the latter study compared 11 conditions instead of only 10, it required more participants to reach the target of 600 simultaneous ratings of each pair of conditions.)
%For the study participants we included, all of them passed all attention checks except for a single participant in the upper body study, who failed one attention check.
%No participant who satisfied these criteria was excluded from the analysis.

Response data from the evaluation and statistical analysis code is provided at \href{https://doi.org/10.5281/zenodo.6939888}{doi.org/10.5281/zenodo.6939888}.

\section{Results and discussion}
% Seems useful for a journal
%The results of the challenge are revolution and a revelation, for the first time finding performance that exceeds the ground-truth data in human-likeness, whilst simultaneously laying bare the true extent of the gap between natural and synthetic motion in terms of speech appropriateness.
%More detail is provided in the sections below.

\subsection{Results of human-likeness studies}
\begin{figure*}[t!]
\centering%
%\hfill
  \begin{subfigure}[b]{0.49\textwidth}
    \centering%
    \includegraphics[width=\textwidth]{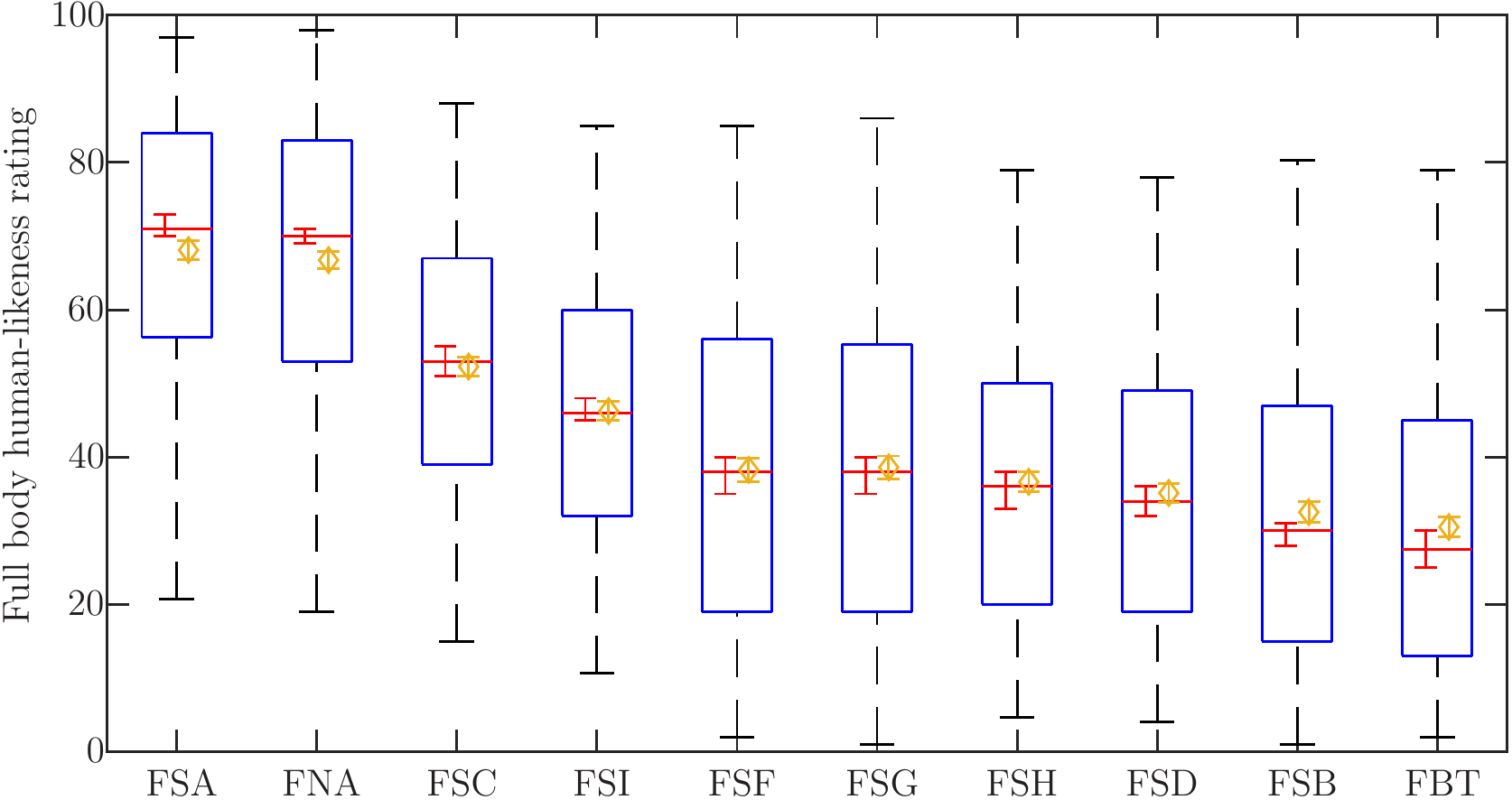}
    \caption{Full-body}
    \label{sfig:fbhumlikeboxplot}
  \end{subfigure}
\hfill
  \begin{subfigure}[b]{0.49\textwidth}
    \centering%
    \includegraphics[width=\textwidth]{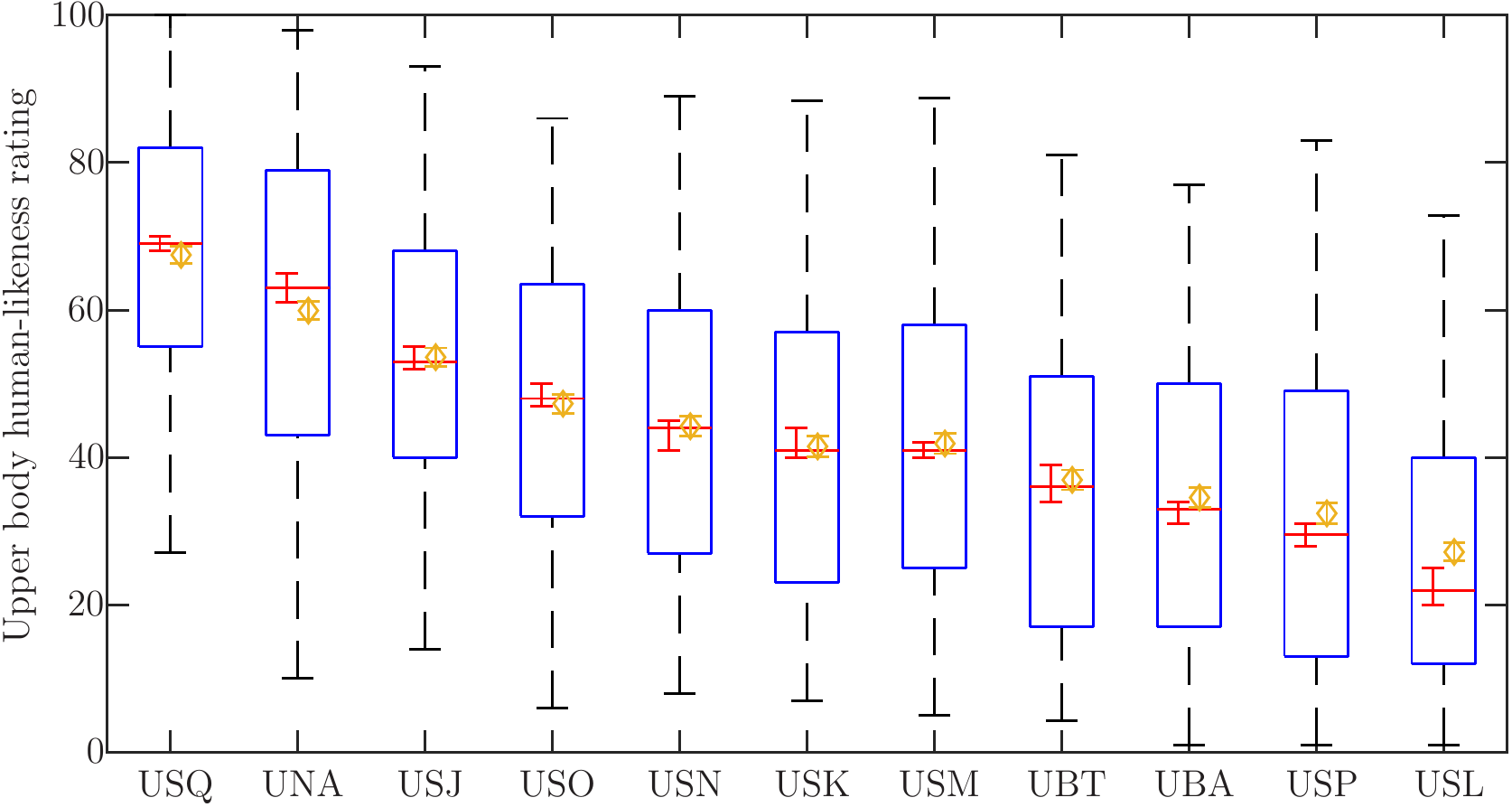}
    \caption{Upper-body}
    \label{sfig:ubhumlikeboxplot}
  \end{subfigure}
%\hfill
\vspace{-2.65pt}
\caption{Box plots visualising the ratings distribution in the human-likeness studies. Red bars are medians and yellow diamonds are means, each with a 0.05 confidence interval and a Gaussian assumption for the means. Box edges are at 25 and 75 percentiles, while whiskers cover 95\% of all ratings for each condition. Conditions are ordered descending by sample median for each tier.}
\label{fig:humlikeboxplots}
\Description{The first box plot shows full-body human-likeness ratings for the 10 conditions. The conditions are sorted in descending order, based on their rating; the order is FSA, FNA, FSC, FSI, FSF, FSG, FSH, FSD, FSB, and FBT. FSA shows a median rating of 71. FBT shows a median rating of 27.5. The second box plot shows upper-body human-likeness ratings for the 11 conditions. The conditions are sorted in descending order, based on their rating; the order is USQ, UNA, USJ, USO, USN, USK, USM, UBT, UBA, USP, and USL. USQ shows a median rating of 69. USL shows a median rating of 22.}
\vspace{-2.65pt}
\end{figure*}

Each test taker in the human-likeness studies contributed 76 ratings to the analyses after removing attention checks, giving a total of 9,196 ratings for the full-body study and 11,400 ratings for the upper-body study.
The results are visualised in Figure\ \ref{fig:humlikeboxplots},
with summary statistics
%(sample median and sample mean)
for the ratings of all conditions
%in each of the two human-likeness studies are shown
given in the first half of Table\ \ref{tab:stats}, together with 95\% confidence intervals for the true median.
%and mean;
These confidence intervals
%for the median
were computed using order statistics, leveraging the binomial distribution cdf; see \cite{hahn1991statistical}.
%while those for the mean used a Gaussian assumption (i.e., using Student's $t$-distribution cdf, rounded outward to ensure sufficient coverage).
%We note that statistics regarding the mean should be interpreted with caution, since responses should be seen as ordinal rather than numerical, and it is therefore improper from a perceptual perspective to perform averaging on the ratings.
\begin{figure*}[t!]
\centering%
%\hfill
  \begin{subfigure}[b]{0.49\textwidth}
    \centering%
    \includegraphics[width=\textwidth]{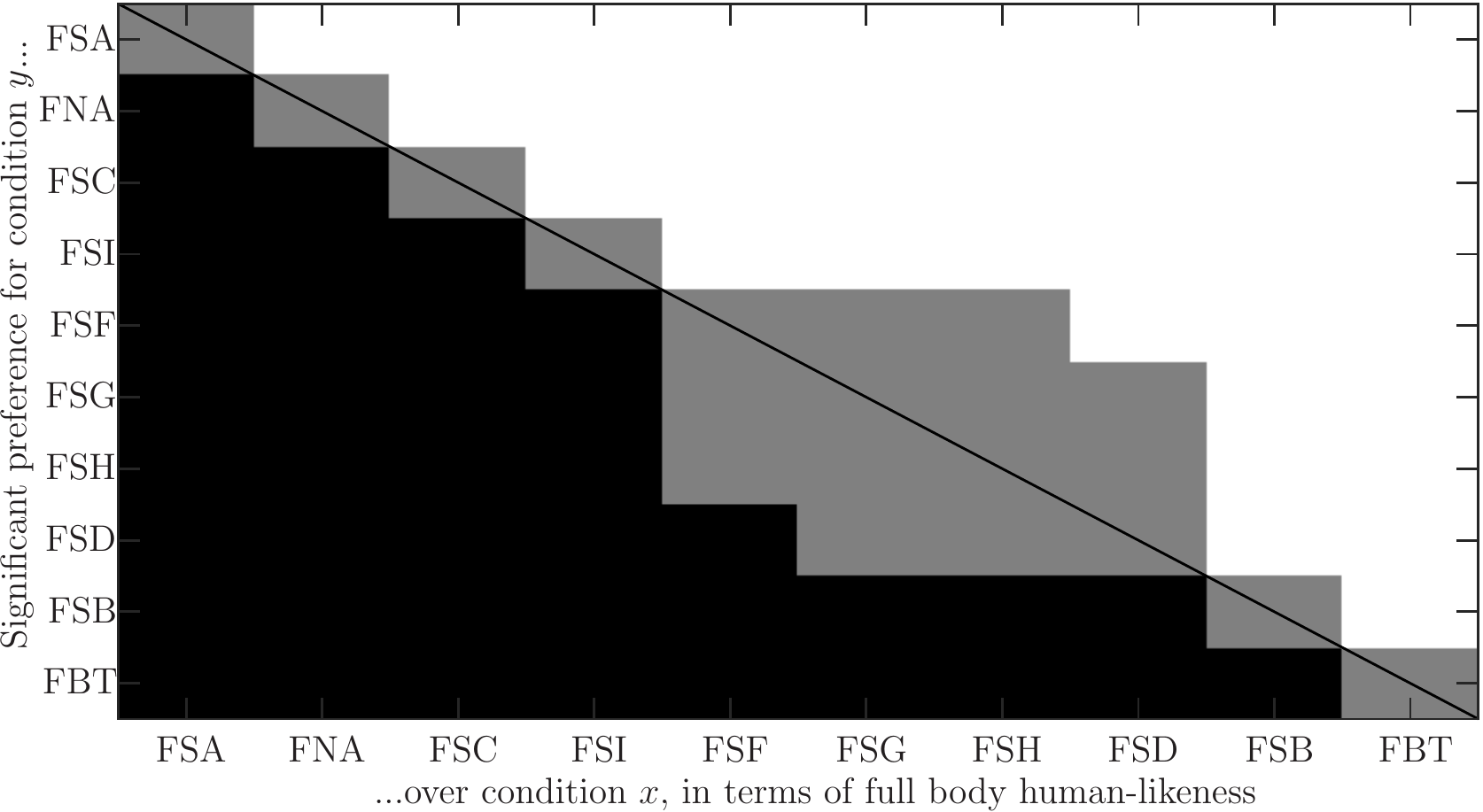}
    \caption{Full-body}
    \label{sfig:fbhumlikedifferences}
  \end{subfigure}
%\hfill
  \begin{subfigure}[b]{0.49\textwidth}
    \centering%
    \includegraphics[width=\textwidth]{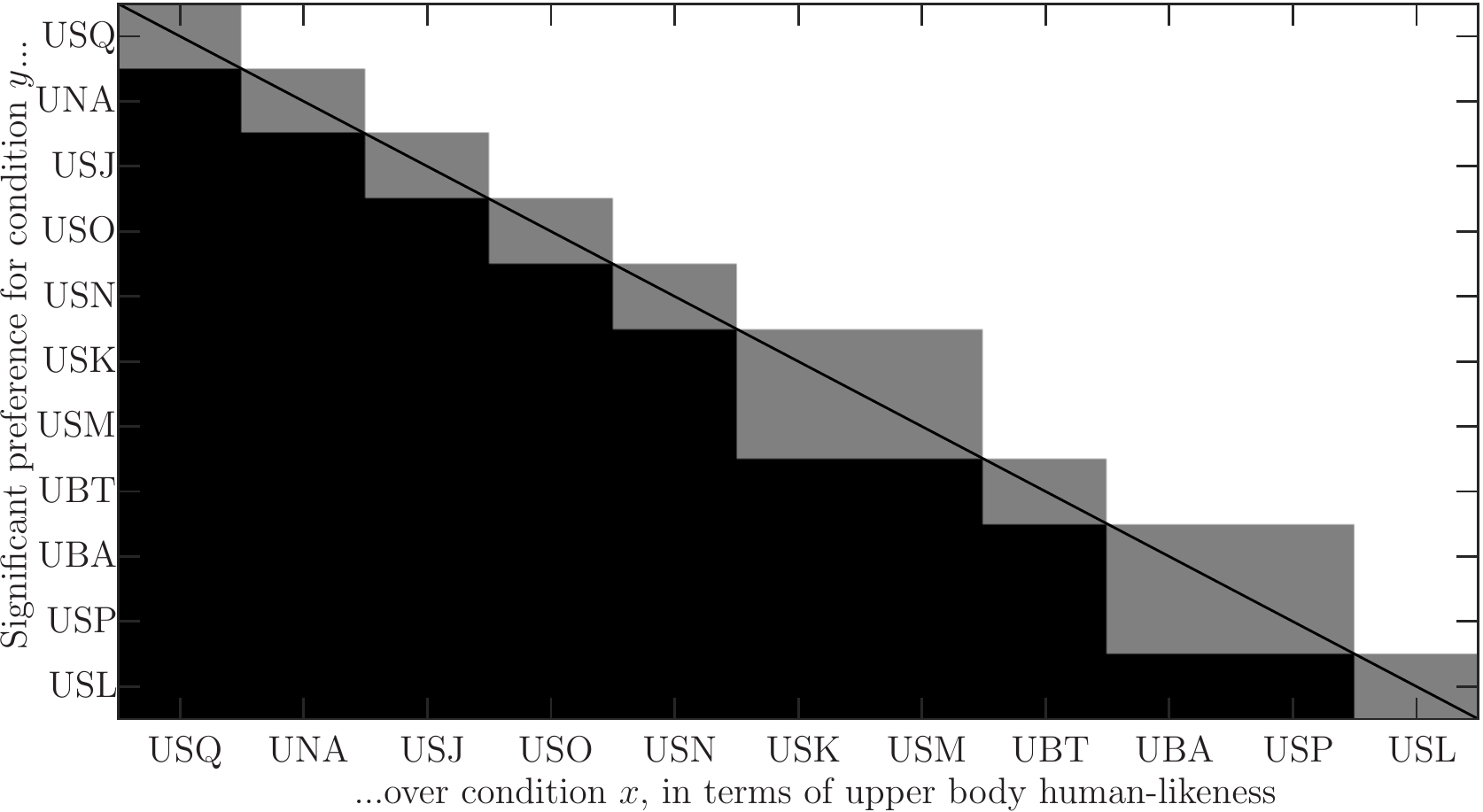}
    \caption{Upper-body}
    \label{sfig:ubhumlikedifferences}
  \end{subfigure}
%\hfill
\vspace{-2.65pt}
\caption{Significant differences in human-likeness. White means the condition listed on the $y$-axis rated significantly above the condition on the $x$-axis, black means the opposite ($y$ rated below $x$), and grey means no statistically significant difference at level $\alpha=0.05$ after Holm-Bonferroni correction. Conditions use the same order as the corresponding subfigure in Figure\ \ref{fig:humlikeboxplots}.}
\label{fig:humlikedifferences}
\Description{These two plots visualize significant differences between every two condition pairs. The first plot shows the results of the full-body human-likeness study. There are 10 conditions, namely: FSA, FNA, FSC, FSI, FSF, FSG, FSH, FSD, FSB, and FBT. FSA, FNA, FSC, FSI, FSB, and FBT showed significantdifferences over all the other conditions. FSF did not show significant preferences over FSG and FSH. FSG did not show significant preferences over FSF, FSH, and FSD. FSH did not show significant preferences over FSF, FSG, and FSD. FSD did not show significant preferences over FSG and FSH. The second plot shows the results of the upper-body human-likeness study. There are 11 conditions: USQ, UNA, USJ, USO, USN, USK, USM, UBT, UBA, USP, and USL. All the conditions except USK, USM, UBA, and USP showed significant differences over all the other conditions. USK did not show significant preferences over USM. USM did not show significant preferences over USK. UBA did not show significant preferences over USP. USP did not show significant preferences over UBA.}
\vspace{-2.65pt}
\end{figure*}

The distributions are seen to be quite broad.
This is common in evaluations like HEMVIP \cite{jonell2021hemvip}, since the range of the responses not only reflects differences between conditions, but also extraneous variation, e.g., between stimuli, in individual preferences, and in how critical different raters are in their judgments.
In contrast, the plotted confidence intervals are seen to be quite narrow, since the statistical analysis can mitigate the effects of much of this variation.

%\subsection{Significant differences}
\label{ssec:pairwisehumlike}
%Despite the wide range of the distributions, the fact that the conditions were rated in parallel on each page enables using pairwise statistical tests to factor out many of the above sources of variation.
%small differences in median rating between different conditions can be teased apart.
To analyse the significance of differences in median rating between different conditions, we applied two-sided pairwise Wilcoxon signed-rank tests to all unordered pairs of distinct conditions in each study.
(This is the same methodology as in the GENEA Challenge 2020 \cite{kucherenko2021large}.)
%This closely follows the analysis methodology used throughout recent Blizzard Challenges and
Unlike Student's $t$-test, which assumes that rating differences follow a Gaussian distribution, this analysis is valid also for ordinal response scales, like those we have here.
For each condition pair, only
cases where both conditions appeared on the same page
%pages for which both conditions were assigned valid ratings
were included in the analysis of significant differences.
%(Recall that not all conditions were rated on all pages due to the limited number of sliders and the presence of attention checks.)
%This meant that every statistical significance test was based on at least 615 pairs of valid ratings in the full-body study, and 603 pairs of valid ratings in the upper-body study.
Because this analysis is based on pairwise statistical tests, it can potentially resolve differences between conditions that are smaller than the width of the confidence intervals for the median in Figure\ \ref{fig:humlikeboxplots}, since those confidence intervals are inflated by variation that the statistical test controls for.
The $p$-values computed in the significance tests were adjusted for multiple comparisons on a per-study basis using the Holm-Bonferroni method \cite{holm1979simple}.%, which is uniformly more powerful than conventional Bonferroni correction.
%to keep the family-wise error rate (FWER), often referred to as alpha-level, at or below $\alpha=0.05$

Our statistical analysis found all but 5 out of 45 condition pairs to be significantly different in the full-body study and all but 2 out of 55 condition pairs to be significantly different in the upper-body study, all at the level $\alpha=0.05$ after Holm-Bonferroni correction.
%Figure\ \ref{fig:appsignificance} visualises the statistically significant differences between conditions using the same condition order as the box plot.
The significant differences we identified
%Which conditions that were found to be rated significantly above or below which other conditions
%in the two studies
are visualised in Figure\ \ref{fig:humlikedifferences}.

\subsection{Discussion of human-likeness results}
\label{ssec:humlikecomments}

%\subsubsection*{On the performance of the best synthetic condition}
Generating convincingly human-like gestures is a difficult problem, and nearly all conditions rated significantly below natural motion capture.
However, each tier contains an entry which is rated significantly above the motion from the motion-capture recordings in terms of human-likeness.
This is a leap forwards from GENEA 2020, and we believe it represents a motion quality not before seen in large-scale evaluations.
That said, we caution that this does not mean that the motion is completely human-like -- indeed, the median rating is much below 100, which would constitute ``completely human-like'' as per our explicit instructions to test takers.
What it does mean is that the motion was perceived as more human-like (in terms of median) than the motion-capture in the database, specifically than the motion-capture data used for FNA/UNA in the subjective evaluation.
In making this distinction, it is important to keep in mind that our human-likeness evaluation is constrained by several factors:
For example, the nominally natural motion is constrained by our ability to accurately capture and visualise human motion.
%, especially the fingers,
%using the technology we used.
Finger motion capture is especially problematic, and the finger motion could not be chosen so as to look completely natural in all segments evaluated, potentially degrading the ratings of FNA/UNA as a result.
Moreover, the use of a deliberately neutral 3D avatar lacking potentially distracting human features such as gaze and lip motion significantly reduces the bandwidth of the communication channel to the user, which lowers the threshold for what needs to be achieved in order to match human motion ratings in the evaluation.
In addition, the greater interquartile range of ratings of UNA compared to FNA could mean that the process of imposing full-body motion from a walking and talking human onto an avatar with fixed lower body may not always yield completely natural results.
An artificial system might have its training data cleaned of problematic instances, so as to prevent it from generating such motion, giving it an edge over UNA.
Future GENEA Challenges intend to only consider full-body motion.

% Seems useful for a journal
\iffalse
\subsubsection*{On the differences between the two tiers}
There are fewer significant differences in the full-body evaluation than in the upper-body evaluation.
Although the difference is not substantial, we would naively expect the opposite, due to the correction for multiple comparisons being more conservative in the upper-body evaluation.
There are many possible explanations for this finding, beyond the fact that the different teams did not all participate in both tiers.
For example, our finding is consistent with an interpretation that full-body motion is a more difficult machine-learning problem, for instance due to increased dimensionality of the output space and the increased number of behaviours that need to be learnt.
This could explain why the best entry in the upper-body evaluation more clearly outperformed UNA, compared to the margin between the best entry in the full-body evaluation and FNA.
%(This would be a competing explanation to the hypothesis in the previous section of artefacts in UNA due to fixing the lower-body to visualise the recorded motion.)
\fi
We found fewer significant differences in the full-body study, perhaps meaning that full-body motion is more difficult to rate consistently.
For example, it contains more behavioural variation, as the character now is moving their legs and changing position,
%in the field of view,
perhaps in response to the conversation partner.
%A contributing factor in this latter case might be that the motion originates from a dyadic conversation, but is evaluated without any contextual information about the interlocutor behaviour, e.g., their pose, stance, or motion with respect to the speaker.
%The absence of such information may be more problematic in assessing the lower-body motion, which might be more closely linked to stance and proxemics in interactions compared to upper-body motion.
Future challenges intend to include information about both conversation parties in the evaluation, so that test takers can be interlocutor-aware.

\subsection{Results of appropriateness studies}
We gathered a total of 8,867 responses for the full-body study and 10,910 responses from the upper-body study that were included in the analysis.
%Every condition received at least 873 responses in the full-body study and 983 in the upper-body study.
\begin{figure*}[t!]
\centering%
%\hfill
  \begin{subfigure}[b]{0.49\textwidth}
    \centering%
    \includegraphics[width=\textwidth]{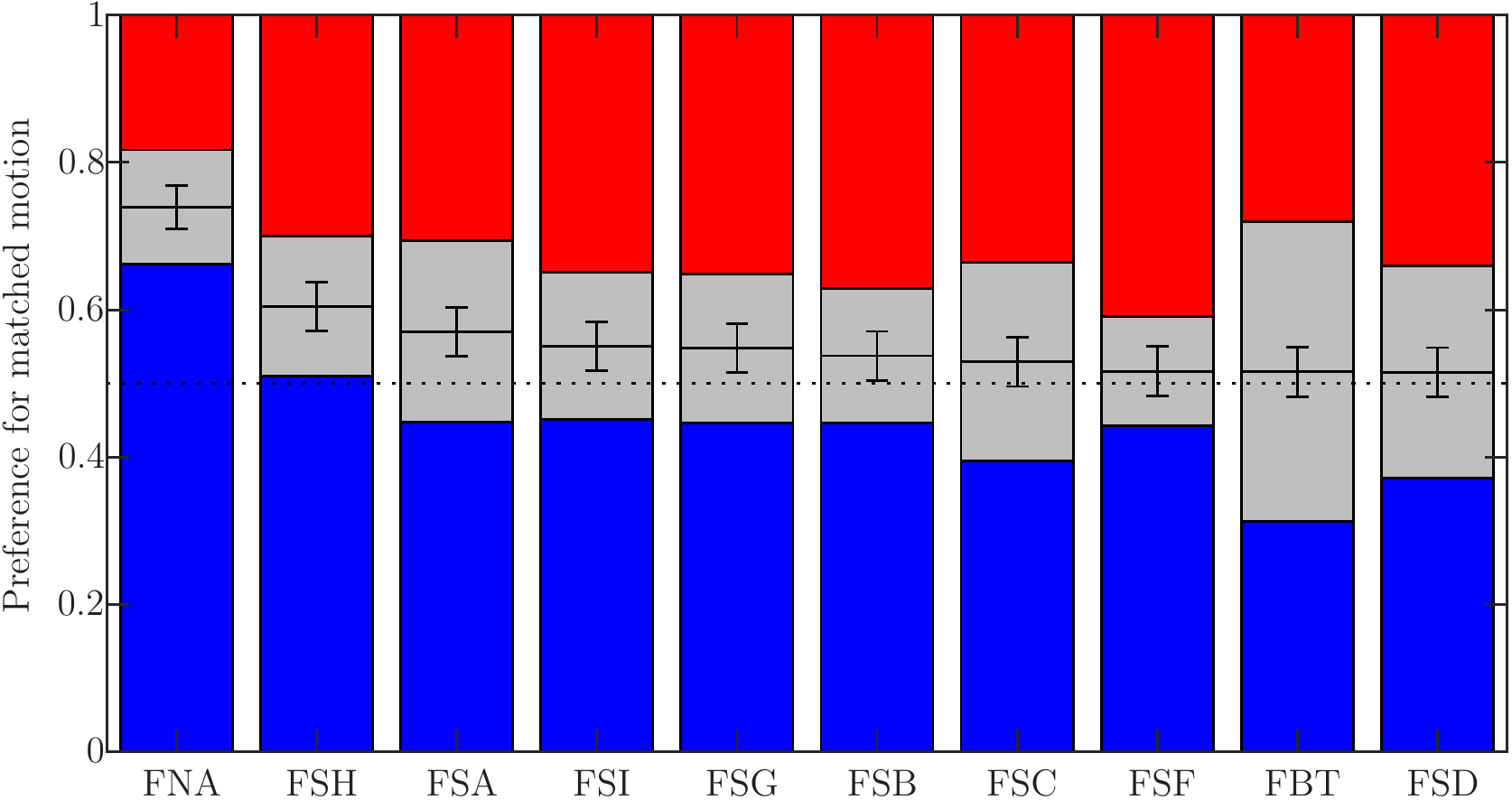}
    \caption{Full-body}
    \label{sfig:fbappropbars}
  \end{subfigure}
\hfill
  \begin{subfigure}[b]{0.49\textwidth}
    \centering%
    \includegraphics[width=\textwidth]{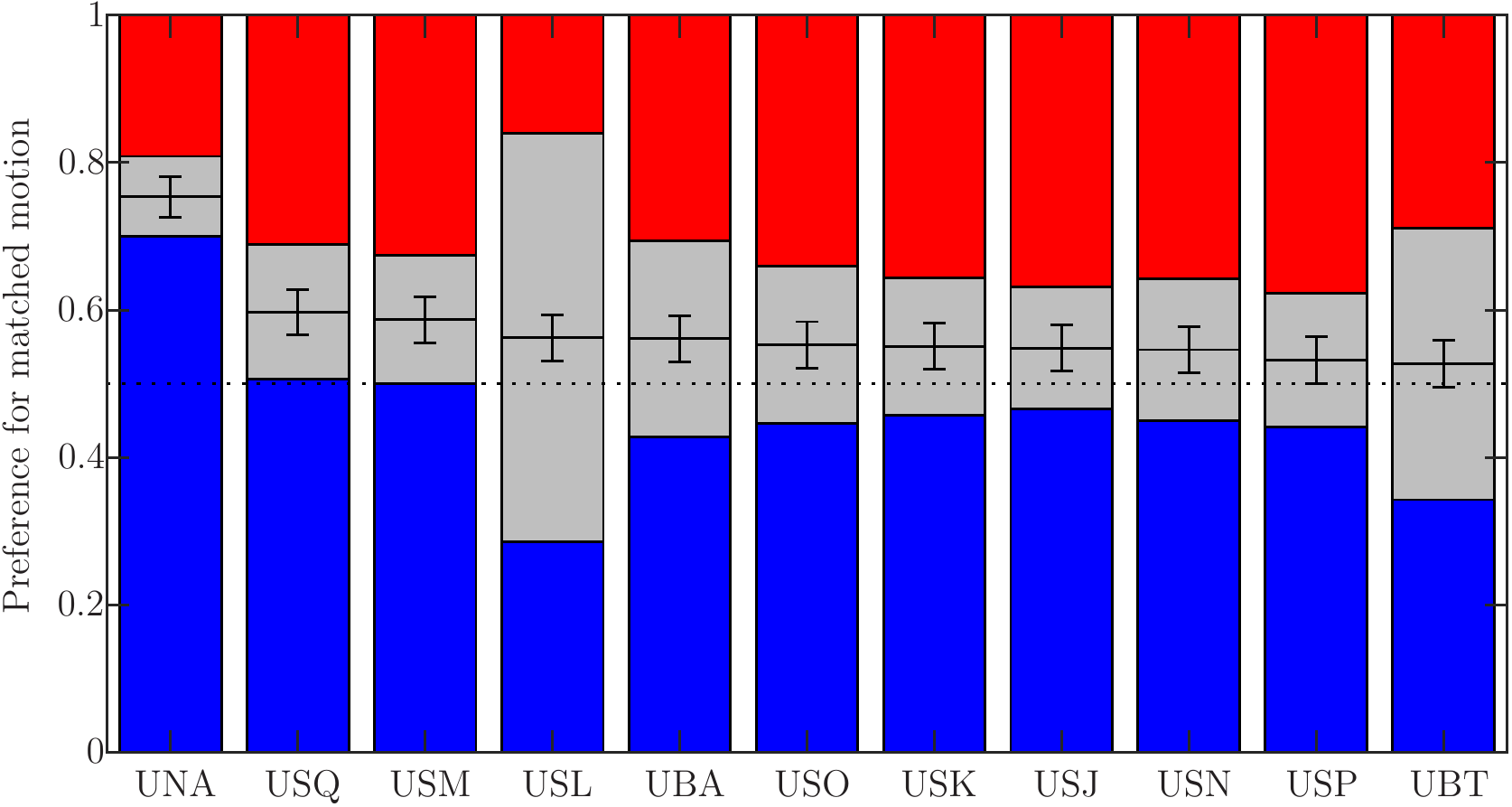}
    \caption{Upper-body}
    \label{sfig:ubappropbars}
  \end{subfigure}
%\hfill
\caption{Bar plots visualising the response distribution in the appropriateness studies. The blue bar (bottom) represents responses where subjects preferred the matched motion, the light grey bar (middle) represents tied (``They are equal'') responses, and the red bar (top) represents responses preferring mismatched motion, with the height of each bar being proportional to the fraction of responses in each category. The black horizontal lines bisecting the light grey bars represent the proportion of matched responses after splitting ties, each with a 0.05 confidence interval. The dotted black line indicates chance-level performance. Conditions are ordered by descending preference for matched motion after splitting ties.}
\label{fig:appropbars}
\Description{The first stacked bar chart shows the preference of matched versus mismatched full-body motion for 10 conditions. The conditions are sorted in descending order, based on the preference for the matching condition. The order is FNA, FSH, FSA, FSI, FSG, FSB, FSC, FSF, FBT, and FSD. For FNA, the blue, gray, and red regions take about 65\%, 15\%, and 20\%, respectively. The second tacked bar chart shows the preference of matched versus mismatched upper-body motion for 11 conditions. The descending order of preference for the matching motion is UNA, USQ, USM, USL, UBA, USO, USK, USJ, USN, USP, and UBT. For UNA, the blue, gray, and red regions take about 70\%, 10\%, and 20\%, respectively.}
\end{figure*}
\begin{figure*}[t!]
\centering%
%\hfill
  \begin{subfigure}[b]{0.49\textwidth}
    \centering%
    \includegraphics[width=\textwidth]{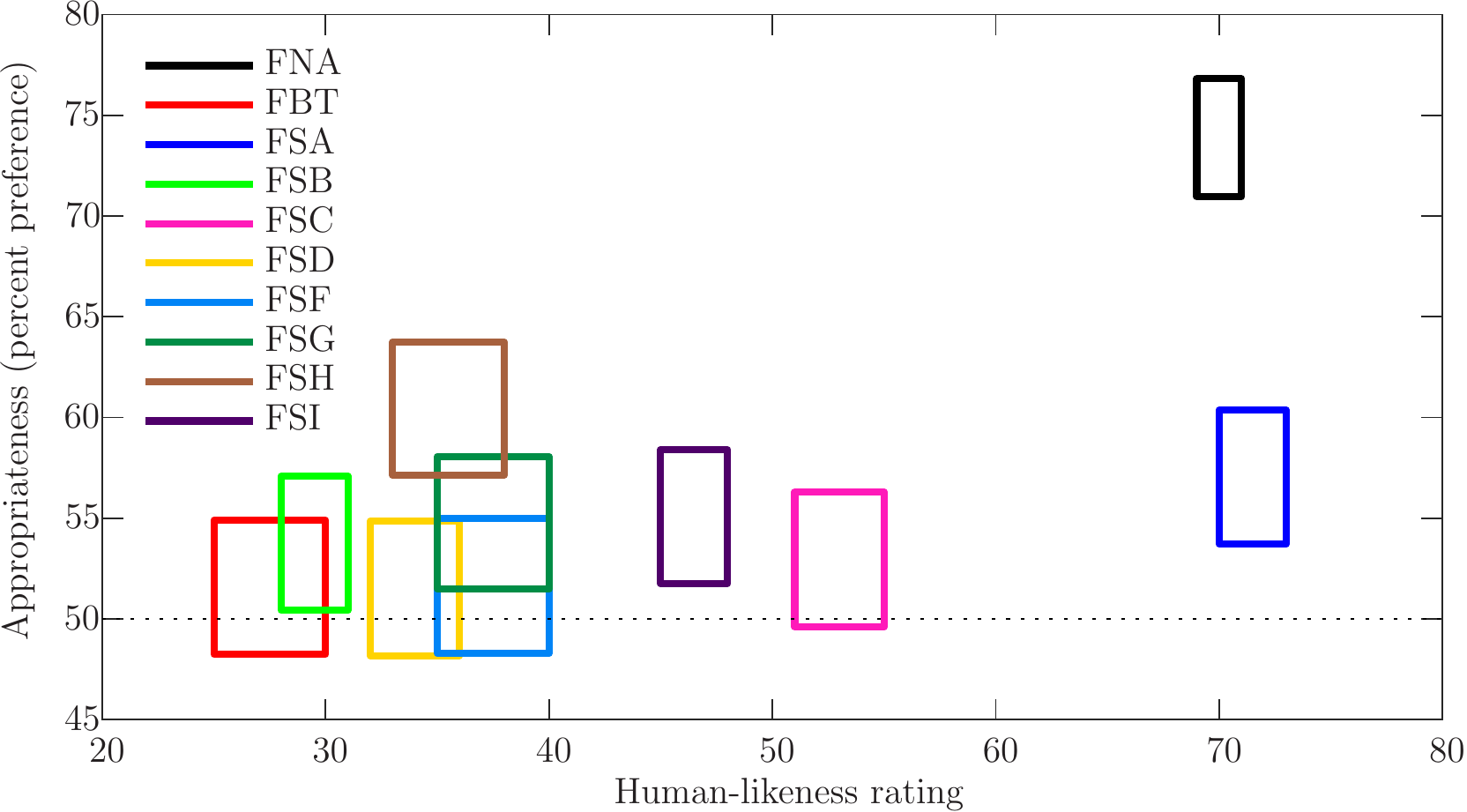}
    \caption{Full-body}
    \label{sfig:fbjpoint}
  \end{subfigure}
\hfill
  \begin{subfigure}[b]{0.49\textwidth}
    \centering%
    \includegraphics[width=\textwidth]{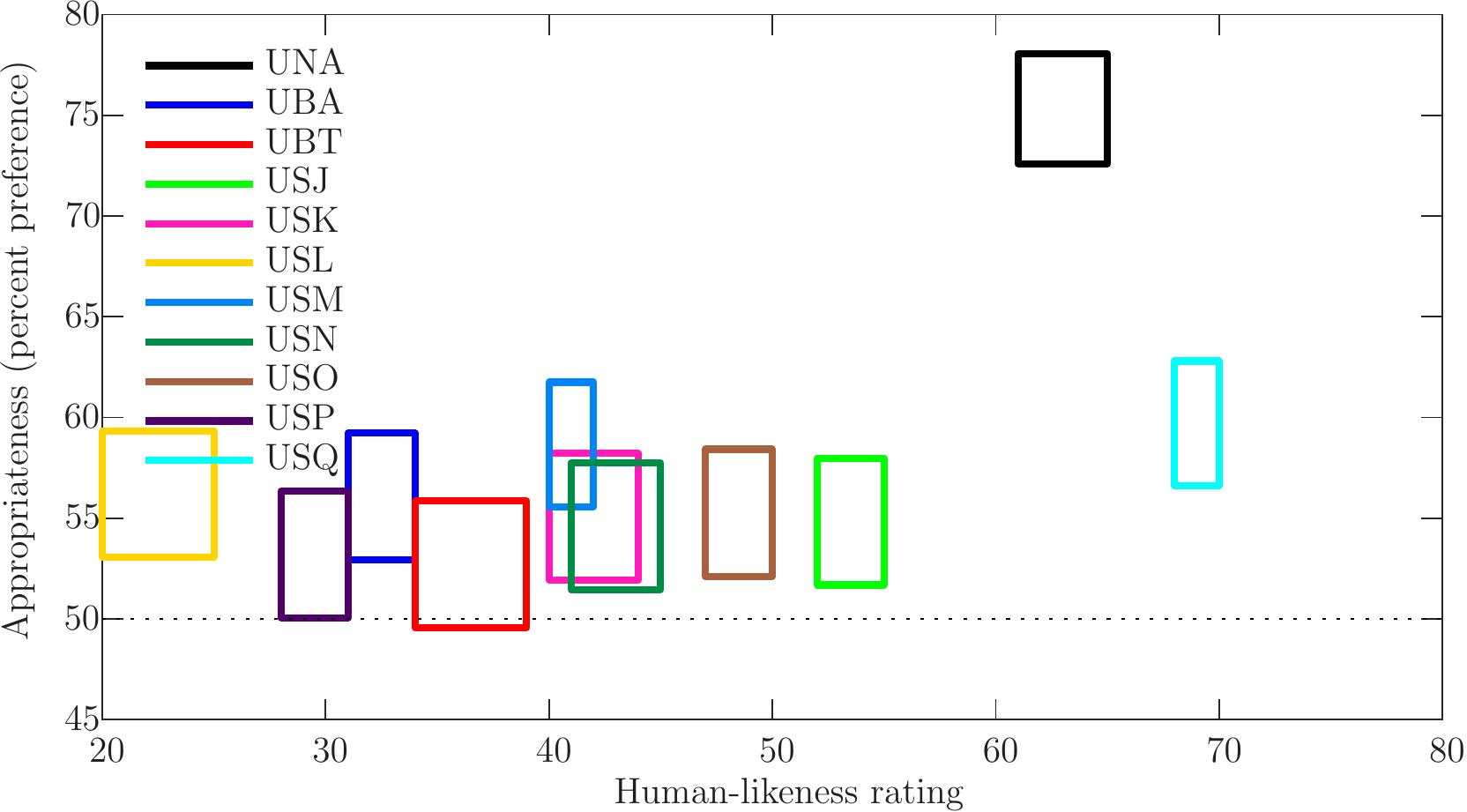}
    \caption{Upper-body}
    \label{sfig:ubjoint}
  \end{subfigure}
%\hfill
\caption{Joint visualisation of the evaluation results for each tier. Box widths show 95\% confidence intervals for the median human-likeness rating and box heights show 95\% confidence intervals for the preference for matched motion in percent, indicating appropriateness.}
\label{fig:joint}
\Description{There are two plots, one for the full-body tier and one for the upper-body tier. In both plots, human-likeness is on the x-axis, and appropriateness (in percentages) on the y-axis.  In the first plot for full-body, the FNA box is in the top right corner, which means high ratings on human-likeness and appropriateness. The FSA box is on the right side. FSC and FSI are on the middle bottom. The other boxes are in the bottom left corner. In the upper-body plot, the UNA box is in the top right corner. The USQ box is on the right side (further right than UNA). The other boxes are in the bottom left corner (around 55 on the y-axis, spread between 20 and 55 on the x-axis).}
\end{figure*}
Raw response statistics for all conditions in each of the two studies are shown in the second half of Table\ \ref{tab:stats}, together with 95\% Clopper-Pearson confidence intervals for the fraction of time that the matched video was preferred over the mismatched, after dividing ties equally between the two groups (rounding up in case of non-integer counts).
The quoted confidence intervals were
%computed using the Clopper-Pearson method and
rounded outward to ensure sufficient coverage.

The response distributions in the two studies are further visualised through bar plots in Figure\ \ref{fig:appropbars}, while Figure\ \ref{fig:joint} visualises the results of the entire challenge in a single coordinate system per tier.
Overall, the distribution of the three different responses across the different conditions is consistent with the mismatching study reported in \cite{jonell2020let}.
No system has a relative preference for matched motion below 50\%, which is the theoretical bottom line, attained by a system whose motion has no relation to the speech.
(Here and forthwith, we only consider the relative preference in the sample after dividing ties equally.)
The greatest relative preference, a 75\% preference for matched motion, is observed for natural motion capture, i.e., FNA/UNA.
%This +25\% effect size over the 50/50 bottom line is about half the theoretical maximum value of +50\% (a 100/0 split).
This should be considered a good result, since previous studies that have incorporated mismatched stimuli, e.g., \cite{jonell2020let,kucherenko2021large}, have found that they sometimes are difficult for participants to distinguish from matched ones, especially if they -- like here -- both correspond to segments where the character is speaking. %(and do not, say, match audio of active speaking with a segment of motion corresponding to the character listening without speaking).
Furthermore, both matched and mismatched motion stimuli have their starting points aligned to the start of a phrase in the speech, meaning that the motion in the stimulus videos might initially be more similar to each other than if the mismatched motion had been excerpted completely at random and not aligned to the start of phrase boundaries.
%Seen in this light, the 75\% preference for matched motion (after splitting ties) in FNA/UNA is a good result.
%The effect size of +25\% is half of the theoretical maximum.
%It is therefore not surprising to find that the preference for matched motion over mismatched motion never is particularly large, even for FNA/UNA.

%\subsection{Significant differences}
Unlike the human-likeness studies, the responses in the appropriateness studies are restricted to three categories and do not necessarily come in pairs for statistical testing in the same way as for the parallel sliders in HEMVIP.
A different method for identifying significant differences therefore needs to be adopted.
We used Barnard's test \cite{barnard1945new} to identify statistically significant differences at the level $\alpha=0.05$ between all pairs of distinct conditions, applying the Holm-Bonferroni method \cite{holm1979simple} to correct for multiple comparisons as before.
%Barnard's test is considered more appropriate than Fisher's exact test for a product of two independent binomial distributions \cite{lydersen2009recommended}, as here.
This analysis found 13 of 45 condition pairs to be significantly different in the full-body study and 10 out of 55 condition pairs to be significantly different in the upper-body study.
%, all at the level $\alpha=0.05$ after Holm-Bonferroni correction.
Specifically, FNA/UNA were significantly more appropriate for the specific speech signal compared to all other, synthetic conditions.
In addition, FSH was significantly more appropriate than FBT, FSC, FSD, and FSF in the full-body study.
No other differences were statistically significant in either study.
%Figure\ \ref{fig:appsignificance} visualises the statistically significant differences between conditions using the same condition order as the box plot.
%Which conditions that were found to be rated significantly above or below which other conditions in the two studies is visualised in Figure\ \ref{fig:humlikedifferences}.
%PDF versions of these images, for optional use in your papers, are provided in the enclosed files\\{}\texttt{full-body\_human-likeness\_median\_pref.pdf} and\\{}\texttt{upper-body\_human-likeness\_median\_pref.pdf} under a Creative Commons Attribution 4.0 International (CC BY 4.0) license.
%This is a much smaller amount of significant differences than we saw in the human-likeness studies.
%For a discussion of these results and how we interpret them, see the next section.

Instead of comparing the appropriateness of different synthesis approaches against one another, one can compare against a random baseline (50/50 performance), and test if the observed effect size is statistically significantly different from zero.
We can assess this at the 0.05 level by checking whether or not the confidence interval on the effect size overlaps with chance performance.
From this perspective, FSA, FSB, FSG, FSH, FSI are significantly more appropriate than chance in the full-body study, and all systems except UBT are more appropriate than chance in the upper-body study.
Unlike other significance tests in this text, these do not include a correction for multiple comparisons.

%Given the fact that the effect size (the difference in preference), while effect sizes between 0 and +10\%.

%\fi

\subsection{Discussion of appropriateness results}
%\subsection{Comments on the results}
%The organisers encourage all teams to think critically and formulate their own take-home messages from the evaluation.
%That said, the organisers would like to share their own perspective on the appropriateness study, and its relationship to other studies here and in GENEA 2020, especially in light of the fact that a new evaluation method is being used.
%wish to make two remarks, one on the performance achieved by the best condition in each study, and one about the differences between the two studies:

%\subsubsection{Summary}
We find the results of the appropriateness evaluation thought-provoking, and revealing about the state of the field.
It is clear that generating meaningful and appropriate gestures is still far from being a solved problem.

We see fewer statistical differences compared to the appropriateness study in GENEA 2020, which asked participants to rate the appropriateness of the stimuli on an absolute scale using HEMVIP \cite{kucherenko2021large}.
However, that study was strongly biased towards conditions with high human-likeness, as discussed in Section\ \ref{sec:approp}.
%evidenced by the fact that mismatched natural motion (M) scored second best in terms of appropriateness there.
In effect, we have traded the high-resolution, high-bias method from GENEA 2020 for a reduced-resolution, low-bias method.
We think this is a step forward, since most prior evaluations of gesture appropriateness for speech have been highly confounded by motion quality, whereas our new methodology is not.
The fact that some synthetic conditions that distinguished themselves the most in terms of appropriateness, namely FSH and USM, exhibited middle-of-the-pack human-likeness, highlights success in disentangling motion appropriateness from motion quality.

%In addition to its strong ability to control for the effect of motion quality, our new method for assessing appropriateness only requires comparing a system to itself. This makes it easy to use and track progress on different sets of stimuli without having to train any baseline systems for the comparisons. This could be advantageous for future benchmarking purposes, since creating appropriate baseline systems is one of the sticking points both for carrying out research and for its subsequent assessment in peer review. Our recommendation for future research is to report effect size and $\alpha=0.05$ Clopper-Pearson confidence intervals similar to Table\ \ref{tab:stats} to easy comparison between studies.

%\subsubsection*{Other trends in the data}
%One other interesting trend in the data is that a few conditions with relatively poor human-likeness, specifically FBT, UBT, and USL, show a noticeably larger fraction of tied responses, compared to other conditions.
%We hypothesise that this could be due to underarticulated motion (after all, a hypothetical, extremely underarticulated system that does not move at all should receive the response ``They are equal'' all the time), but we have not yet looked closely at the actual stimuli to verify or disprove this hypothesis.

\section{Conclusions and implications}
%Taken together with the naturalness evaluation, these results illustrate that we currently have the means to be able to generate quite convincing data-driven 3D gesture motion (although attaining that level of quality is something that only few systems are capable of at present), but we are only at the beginning of the road when it comes to generating co-speech motion that is appropriate for the specific speech.

We have hosted the GENEA Challenge 2022, to directly compare many different gesture-generation methods and assess the state of the art in data-driven co-speech gesture generation.
%for full-body and upper-body avatars.
%The central design goal of the challenge was to enable direct comparison between many different gesture-generation methods while controlling for factors of variation external to the model, namely data, embodiment, and evaluation methodology, and to disentangle the effects of motion human-likeness and motion appropriateness in the evaluations.
Our evaluation results show that, with the right method, synthetic motion can attain a human-likeness equal or better than the underlying motion-capture data.
This is a big step forwards compared to the 2020 challenge.
However, using a new evaluation paradigm, we find that synthetic gestures are much less appropriate for the speech than human gestures, also when controlling for differences in human-likeness.
%between the different conditions.
%suggest that the field is advancing measurably, since most submissions performed significantly better than the baselines.
%Different systems were also found to be good at different things on the two scales (human-likeness and appropriateness) that we assessed.
%However, a substantial gap remains between synthetic and natural gesture motion, indicating that gesture generation is far from a solved problem.

%The challenge findings have implications for both research and practice, and we attempt to summarise our perspective on those here:

% Seems useful for a journal
%\subsubsection*{Practical systems}
%If you are building a gesture-generation system, you can reach near top-of-the-line human-likeness and an appropriateness not far behind most synthetic systems, simply by relying on playback of pre-recorded gestures, without much (or any) regard for the speech beyond its on and offset. This would save a lot of technical complexity.

%\subsubsection*{Research systems}
%If you are are working on or with gesture generation in your research, .

We believe the challenge adds value to the research community in many ways.
A lot can doubtlessly be learnt from the system-description papers by the participating teams.
The materials we release from the challenge (e.g., time-aligned splits of audio, text, and gesture data;
visualisation; code; and evaluation stimuli and responses) can have broad use for future benchmarking and research in gesture generation, similar to what happened after the 2020 challenge.
%
%\subsubsection*{Evaluations}
%For evaluations and evaluation research...
%
In particular, the new methodology we demonstrate for assessing motion appropriateness for speech is much more accurate at controlling for the effect of motion quality and
%importantly,
does not involve subjects making any direct comparisons between videos generated by different conditions.
%This may mean that results (effect sizes with confidence intervals) can likely be directly compared between different studies on the same data, at least as long as the same speech segments are evaluated, \emph{without} having to include the various other synthetic baseline conditions in the new evaluation.
We believe this may enable direct comparison between different studies on the same data, \emph{without} having to include the various other synthetic baseline conditions in the new user study.
%This is a big simplification compared to parallel methodologies like HEMVIP \cite{jonell2021hemvip}.
%, involve simultaneously comparing and evaluating many different conditions against each other.
%, since responses in those studies are affected by what other videos are shown on the same page, and studies thus cannot be directly compared unless stimuli or implementations of previous synthetic baseline conditions are included in the new study.

%\subsubsection*{Expected future developments}
Based on the fact that one condition in each tier managed to achieve excellent human-likeness, we expect that, in the medium-term future, gesture-generation systems should be able to advance
to more consistently match
%so as to more consistently match or possibly even exceed
motion capture in terms of human-likeness.
This is similar to recent developments in verbal behaviour generation, where neural language models \cite{brown2020language} and speech synthesisers \cite{shen2018natural,li2019neural} trained on large datasets are approaching the text and speech produced by humans in terms of surface quality (but not necessarily appropriateness).
%Gesture generation may be lagging behind due to relative scarcity of motion-capture data, compared to text and audio.
As that evolution runs its course, 
%Over time, however,
we believe that research into appropriate rather than human-like motion is poised to become the new frontier in gesture generation.
There is already evidence that existing deep-learning methods in principle can predict even the hard case of semantically motivated, communicative gestures from speech \cite{kucherenko2022multimodal, kucherenko2021speech2properties2gestures}.

%\subsubsection*{Future challenges}
%Implications for future challenges
We think that future challenges should study more difficult scenarios that are farther from being solved, for example full-body motion in dyadic interaction.
That can also provide interesting opportunities for exploring other types of appropriateness, e.g., with respect to the interlocutor stance and behaviour, as studied in \cite{jonell2020let}.
%Generating interlocutor-aware full-body gestures will therefore be a focus of GENEA 2023.
%Furthermore, we think
In general, challenges like the one described here can play an important part in identifying key factors for generating convincing co-speech gestures in practice, and help drive and validate future progress toward the goal of endowing embodied agents with natural and appropriate gesture motion.

\begin{acks}
%The GENEA Challenge 2022 used the Talking With Hands Dataset collected by Facebook. The challenge dataset was further processed by Taras Kucherenko, Teodor Nikolov, Mihail Tsakow.

The authors wish to thank Meta Research for the data; Carolyn Saund, Axel Johansson, Christianne Sandstig, Leonhard Grosse, Natalia Kalyva, and Natasha Greenwood for the transcriptions; Esther Ericsson for the 3D character; Zerrin Yumak for input; Judith Bütepage, Minsu Jang, and Tony Belpaeme %Andr{\'e} Tiago Abelho Pereira, Bram Willemsen, Dmytro Kalpakchi, Jonas Beskow, Kevin El Haddad, and Ulme Wennberg
for informal review.
%feedback on the paper preprint.

This research was partially supported by IITP grant no.\ 2017-0-00162 (Development of Human-care Robot Technology for Aging Society) funded by the Korean government (MSIT), by the Flemish Research Foundation (FWO) grant no.\ 1S95020N, by the Portuguese Foundation for Science and Technology grant no.\ SFRH/BD/127842/ 2016, and by the Knut and Alice Wallenberg Foundation, both through Wallenberg Research Arena (WARA) Media and Language -- with in-kind contribution from the Electronic Arts (EA) R\&D department, SEED -- and through the Wallenberg AI, Autonomous Systems and Software Program (WASP).
\end{acks}

\bibliographystyle{ACM-Reference-Format}
\bibliography{refs}


\end{document}